\documentclass[paper]{JHEP3}
\usepackage{cite}
\usepackage{amssymb,amsmath}
\usepackage{graphicx}

\def\bom#1{{\mbox{\boldmath $#1$}}}

\def\Q2{\left(Q^{2}\right)}
\def\e{\epsilon}

\def\CA{C_A}

\def\NF{N_F}
\def\NFZ{N_{F,V}}
\def\MSbar{$\overline{{\rm MS}}$}

\def\l({\left(}
\def\r){\right)}

\def\e{\epsilon}

\def\sab{s_{12}}
\def\sac{s_{13}}
\def\sbc{s_{23}}

\def\sabc{s_{123}}

\newcommand{\p}{p\hspace{-1.1ex}/}
\newcommand{\ep}{\epsilon\hspace{-1.0ex}/}

\title{Two-loop QCD helicity amplitudes for $q\bar q\to W^\pm\gamma$ and 
$q\bar q \to Z^0\gamma$}
\author{Thomas Gehrmann, Lorenzo Tancredi\\Institut f\"ur Theoretische Physik, 
Universit\"at Z\"urich, Wintherturerstrasse 190,\\CH-8057 Z\"urich, Switzerland}

\keywords{QCD, Collider Physics, NLO and NNLO Calculations}
\abstract{The self-couplings of the electroweak gauge bosons are probed at hadron 
colliders through the production of a massive gauge boson and a photon. To extend the 
theoretical description of this type of final states towards next-to-next-to-leading order (NNLO)
in QCD, 
we derive the two-loop QCD corrections to the helicity 
amplitudes describing the production of a massive gauge boson in association 
with a real photon. Our results are obtained by applying projectors to the general 
parton-level tensor structure. The leptonic decay of the vector boson is included, thus allowing 
for a fully exclusive description of the final state. The infrared poles of the amplitudes 
are described by an infrared factorization formula. We provide an analytic expression for the 
finite remainder of the amplitude in terms of one- and two-dimensional harmonic 
polylogarithms. The amplitudes are expressed in the physical kinematics relevant to 
gauge-boson-plus-photon production at hadron colliders. As a by-product, we also derive 
the two-loop QCD amplitudes for vector-boson-plus-jet production at hadron colliders.  }
\preprint{{ZU-TH 26/11}}

\begin{document}
\allowdisplaybreaks

\section{Introduction}
Pair production of electroweak gauge bosons ($\gamma, W^\pm, Z^0$) 
 offers a wide spectrum of observables, 
which allow to test the theory of the electroweak interaction, to 
probe the Higgs mechanism of electroweak symmetry breaking 
and to search for physics beyond the 
standard model.  While photons are directly observed in the 
detector, the massive $W$ and 
$Z$ bosons are identified from their leptonic decay modes. 

The standard model predicts specific values 
and structures for the  couplings among the electroweak 
gauge bosons: $W^\pm$, $Z^0$ and $\gamma$. Physics effects beyond the 
standard model could modify these gauge boson 
self-couplings~\cite{ancoupl1,ancoupl2}. 
Observations 
of such anomalous couplings may help to constrain new theory models and 
could  provide 
indirect evidence for new physics effects at energy scales  
above the nominal collision energy. The couplings of 
the massive $W^\pm,Z^0$ bosons to the photon are determined 
at hadron colliders by measuring $W^\pm\gamma$
and $Z^0\gamma$ production cross sections and comparing them to 
theoretical predictions. Measurements have been carried out at 
Tevatron~\cite{tev},
and first results from the LHC are already becoming available~\cite{lhc}. 
The accuracy of the coupling determination is 
potentially limited by both the experimental accuracy and by 
uncertainties inherent to the theoretical prediction. 

At present, $W^\pm\gamma$
and $Z^0\gamma$ production at hadron colliders is described theoretically 
to next-to-leading order (NLO) in QCD~\cite{nlovgamma,ancoupl2,dixon2} and to NLO 
in the electroweak theory~\cite{nloew}. With increasing order 
in perturbative QCD, new production channels (with new combinations 
of parton distributions) for vector boson pairs 
start contributing; the complete spectrum of partonic channels is 
only present from next-to-next-to-leading order (NNLO) 
onwards. 
Moreover, the inclusion of 
NNLO corrections to gauge boson pair production will lower the 
inherent theoretical uncertainty of the prediction (usually quantified 
by variation of the renormalisation and factorisation scales) and allow 
for a fully consistent inclusion of NNLO parton distribution 
functions. 

The calculation of gauge boson pair production at NNLO requires three 
types of ingredients: the two-loop partonic $2\to 2$ matrix elements for 
the production of the gauge boson pair under consideration, the one-loop 
partonic $2\to 3$ matrix elements for the production of the gauge boson 
pair in association with an extra parton and the tree-level $2\to 4$ 
matrix elements involving two extra partons. The latter two contributions 
are equally contributing to the NLO corrections for the production of 
a vector boson pair with an extra jet, which have been computed for 
$\gamma\gamma j$~\cite{ggjnlo}, $V\gamma j$~\cite{vgjnlo} and 
$VVj$~\cite{vvjnlo} already some time ago. At NNLO, the 
contributions from both these channels will contain infrared 
singularities from one or two final state partons becoming soft or collinear. 
These singularities cancel only when combined with the 
infrared-singular two-loop contributions, such that a method is needed 
for their extraction from the real radiation processes. 
Several methods have been applied successfully in 
NNLO
calculations of exclusive observables in the recent past: 
sector decomposition~\cite{secdec}, $q_T$-subtraction~\cite{qtsub} and 
antenna subtraction~\cite{antsub,hadant,joao}. It should be noted 
that the $q_T$-subtraction method is restricted to observables that 
are described by non-QCD processes at leading order, which is the case 
for vector boson pair production. The first 
calculation of NNLO corrections to a vector boson pair production 
process ($pp\to \gamma\gamma$) has been performed most recently~\cite{ggnnlo}
using this method. 

The two-loop parton-level matrix elements for vector boson pair 
production are at 
present known only for $q\bar q \to \gamma \gamma$~\cite{qqgg2l} 
and $gg\to \gamma\gamma$~\cite{gggg2l} (where the latter 
two-loop amplitude formally 
contributes only beyond NNLO). The high energy approximation for 
$q\bar q \to W^+W^-$ and $q\bar q \to Z^0Z^0$ has also been 
derived~\cite{greg}. It is the purpose of this paper to compute the 
two-loop corrections to the matrix elements for the production of 
a massive vector boson and a photon: $q\bar q \to W^\pm\gamma$ and $q\bar q 
\to Z^0\gamma$. The calculation follows closely the techniques that were 
employed in the calculation of two-loop corrections to the 
$\gamma^*\to q\bar q g$ matrix elements~\cite{3jme,3jtensor}, which 
were a crucial ingredient to the NNLO corrections~\cite{ourevent,weinzierl}
to three-jet production 
and related event shapes
at $e^+e^-$ colliders.  

This paper is structured as follows: in Section~\ref{sec:kin}, we 
fix the notation and discuss the basic helicity structure of the process under
consideration. The calculation of the two-loop amplitudes is 
described in Section~\ref{sec:calc}, and the results are discussed 
in Section~\ref{sec:results}. The two-loop helicity amplitudes are 
obtained in a closed analytic form, which is however too large to be 
quoted in the paper, and we enclose computer algebra files containing 
the results with the submission. 
We performed several non-trivial checks on the results,
which are described in Section~\ref{sec:checks}. We conclude with an 
outlook in Section~\ref{sec:conc}.
 We enclose appendices with the one-loop 
helicity amplitudes, examples of selected colour coefficients in the two-loop 
amplitudes and with a discussion on kinematical crossings, including the 
two-loop matrix elements relevant to vector-boson-plus-jet production:
$q\bar q \to V g$ and $qg \to Vq$. 

\section{Kinematics and basic helicity structure}
\label{sec:kin}
The production of a massive vector boson and a photon in quark-antiquark annihilation is 
related by crossing to the decay of the vector boson into a quark-antiquark-photon 
final state, which has the same kinematics as three-jet-production $(3j)$ in $e^+e^-$ 
annihilation. Technically, the calculation of QCD corrections to the $q\bar q\to V\gamma$ 
amplitudes is thus similar to previous calculations for the helicity amplitudes for 
$3j$-production, which have been derived to two-loop accuracy in QCD~\cite{3jtensor}. 

Including the leptonic decay of the vector boson, the partonic subprocesses 
yielding $V\gamma$ final states are:
\begin{eqnarray*}
&&q(p_2) + \bar q(p_1) \to \gamma(-p_3) + Z^0(q) 
\to \gamma(-p_3) + l^+(p_5) + l^-(p_6)\;,\\
&&q(p_2) + \bar q'(p_1) \to \gamma(-p_3) + W^-(q) 
\to \gamma(-p_3) + \bar \nu(p_5) + l^-(p_6)\;,\\
&&q(p_2) + \bar q'(p_1) \to \gamma(-p_3) + W^+(q) 
\to \gamma(-p_3) +  l^+ (p_5) + \nu (p_6)\;.
\end{eqnarray*}
The $Z^0$-boson process implicitly includes also a contribution from an
off-shell photon $\gamma^*$.
In the most general case where two quarks of two different flavours appear, 
to fix the conventions, we will refer from now on to  $p_1$ as the
momentum of the anti-quark $\bar{q}'$ and as $p_2$ to the momentum of the 
quark $q$.
The momentum of the vector boson is given by
\begin{equation}
q^{\mu} = p_5^{\mu}+p_6^{\mu}\; .
\end{equation}

It is convenient to define the invariants
\begin{equation}
\sab = (p_1+p_2)^2\;, \qquad \sac = (p_1+p_3)^2\;, \qquad 
\sbc = (p_2+p_3)^2\;,
\end{equation}
which fulfil
\begin{equation}
q^2  =(p_1+p_2+p_3)^2 = \sab + \sac + \sbc \equiv \sabc \; ,
\end{equation}
as well as the dimensionless invariants
\begin{equation}
x = \sab/\sabc\;, \qquad y = \sac/\sabc\;, \qquad z = \sbc/\sabc\;,
\end{equation}
which satisfy $x+y+z=1$.

In $3j$-production, $q^2$ is time-like (hence 
positive) and all the $s_{ij}$ are also positive, which implies that 
$x,y,z$ all lie in the interval $[0;1]$, with the above constraint $x+y+z=1$.
For the $V\gamma$ production, $q^2$ remains time-like, but only $s_{12}$
is positive:
\begin{equation}
q^2 > 0\;,\quad s_{12} > 0\;,\quad s_{13} < 0, \quad s_{23} < 0\; ,
\end{equation}
or, equivalently,
\begin{equation}
x>0\;, \quad y<0 \;, \quad z<0\; .
\end{equation}
It was shown in~\cite{ancont} that the kinematical situation of this 
configuration can be expressed by introducing new dimensionless 
variables
\begin{equation} 
u = -\frac{s_{13}}{s_{12}}=-\frac{y}{x}\,, \qquad v = \frac{q^2}{s_{12}} =
 \frac{1}{x}\;,
\end{equation}
which fulfil
\begin{displaymath}
0\leq u \leq v\,, \qquad 0\leq v \leq 1\;.
\end{displaymath}

The helicity amplitudes for $q\bar q\to V\gamma$ can be expressed as 
a product of a partonic current $S_\mu$ and a leptonic current $L_\mu$:
\begin{equation}
 A(p_5,p_6;p_1,p_3,p_2) = L^\mu(p_5;p_6) S_\mu(p_1;p_3;p_2)\;.
 \end{equation}
Only the partonic current receives contributions from QCD radiative corrections,
and it can be perturbatively decomposed as:
\begin{align}
 S_\mu(p_1;p_3;p_2) = 
\sqrt{4 \pi \alpha} \, 
&\Big(\, S^{(0)}_\mu(p_1;p_3;p_2) 
+ \left(\frac{\alpha_s}{2\pi}\right)  S^{(1)}_\mu(p_1;p_3;p_2)\nonumber \\
&+ \left(\frac{\alpha_s}{2\pi}\right)^2 S^{(2)}_\mu(p_1;p_3;p_2) 
+ {\cal O}(\alpha_s^3) \, \Big).
\end{align}
It is a colour-singlet.
The vector boson decay to a lepton-antilepton pair is described by a 
leptonic current. 
To be as general as possible, we consider only the basic amplitude structure in 
the partonic and leptonic current, and include charges and coupling factors 
related to the massive vector boson
only when assembling the final results.  We have extracted a factor 
$e = \sqrt{4 \pi \alpha}$ for the photon coupling in the partonic current, such that 
all the quark charges will be expressed in units of $e$.

The most general structure of the partonic current can be derived 
from symmetry considerations~\cite{3jtensor}:
\begin{align}
S_{\mu}(p_1;p_3;p_2) &= A_{11}\; T_{11 \mu} + A_{12}\; T_{12 \mu} +  A_{13}\; T_{13 \mu}
\nonumber \\
&+ A_{21}\; T_{21 \mu} + A_{22}\; T_{22 \mu} +  A_{23}\; T_{23 \mu}\nonumber \\
&+ B\;T_{\mu}\;, \label{eq:smubase}
\end{align}
where $T_{ij \mu}$ and $T_{\mu}$ are the following tensor structures:
\begin{align}
&T_{1j \mu} = \bar{v}(p_1) \left[ \p_3\; \epsilon_3 \cdot p_1 \; p_{j \mu} - {s_{13} \over 2} \ep_3 \; p_{j\mu}
                                + {s_{j4}\over 4} \ep_3 \p_3 \gamma_{\mu} \right] u(p_2),\nonumber \\
&T_{2j \mu} = \bar{v}(p_1) \left[ \p_3\; \epsilon_3 \cdot p_2 \; p_{j \mu} - {s_{23} \over 2} \ep_3 \; p_{j\mu}
                                + {s_{j4}\over 4} \gamma_{\mu} \p_3 \ep_3 \right] u(p_2),\nonumber \\
&T_{\mu} \;\;\; = \bar{v}(p_1) \left[ s_{23} \left( \gamma_{\mu} \epsilon_3 \cdot p_1 + {1 \over 2} \ep_3 \p_3 \gamma_{\mu} \right)
                              -s_{13} \left( \gamma_{\mu} \epsilon_3 \cdot p_2 + {1 \over 2} \gamma_{\mu} \p_3 \ep_3  \right) \right] u(p_2),
\end{align}
where we defined: 
\begin{equation}
 s_{14} = s_{12} + s_{13},\quad 
 s_{24} = s_{12} + s_{23},\quad
 s_{34} = s_{13} + s_{23}. \nonumber  
\end{equation}

The tensor coefficients $A_{ij}$ and $B$ can be determined by 
appropriate projectors, applied to the Feynman-diagrammatic expression 
of the amplitude. Projections on the diagrams are performed in 
dimensional regularisation in $d=4-2\e$ dimensions. The projectors can be found 
in~\cite{3jtensor}. \newline
Each of the unrenormalised
coefficients $A_{ij}$ and $B$ has a perturbative expansion of the form
\begin{eqnarray}
A_{ij}^{{\rm un}} &=& 
\sqrt{4 \pi \alpha} \,\left[
A_{ij}^{(0),{\rm un}}  
+ \left(\frac{\alpha_s}{2\pi}\right) A_{ij}^{(1),{\rm un}}  
+ \left(\frac{\alpha_s}{2\pi}\right)^2 A_{ij}^{(2),{\rm un}} 
+ {\cal O}(\alpha_s^3) \right] \;,\nonumber \\
B^{{\rm un}} &=& 
\sqrt{4 \pi \alpha} \,  \left[
B^{(0),{\rm un}}  
+ \left(\frac{\alpha_s}{2\pi}\right) B^{(1),{\rm un}}  
+ \left(\frac{\alpha_s}{2\pi}\right)^2 B^{(2),{\rm un}} 
+ {\cal O}(\alpha_s^3) \right], \;
\end{eqnarray}
where the dependence on $(\sac,\sbc,\sabc)$ is implicit.

By fixing the helicities of the partons, the partonic current can be
cast in the usual spinor helicity notation~\cite{dixon}. 
All helicity configurations can be obtained from the amplitude
\begin{align}
S_R^\mu&(p_1^-;p_3^+;p_2^+) = {1 \over \sqrt{2}} 
\langle 1 2\rangle [1 3]^2 
\left( p_{1 \mu} A_{11} + p_{2 \mu} A_{12} + p_{3 \mu} A_{13} \right) 
- {1 \over \sqrt{2}}  {\langle1 2\rangle [1 3] \over \langle 2 3\rangle } 
[1\;| \gamma_\mu |\;2 \rangle \; s_{23} B \notag \\
&+ {1 \over \sqrt{2}}
[1 3] [3\;| \gamma_\mu|\;2\rangle \; 
\left[ s_{23}B + {1 \over 2} \left( (A_{11}+A_{12})s_{12} + (A_{11}+A_{13})s_{13} + (A_{12}+A_{13})s_{23} \right) \right] \label{RightS}
\end{align}
by charge and parity conversion. 
For $\bar{q}'(p_1)$, $q(p_2)$ incoming, the above amplitude 
corresponds to a right-handed current.
Notice that \eqref{RightS} has been obtained
assuming that the momentum of the photon is $-p_3$.
We have:
\begin{align}
S_L^\mu(p_1^+;p_3^-;p_2^-) &= \left[S_R^\mu(p_1^-;p_3^+;p_2^+)\right]^*, \nonumber \\
S_L^\mu(p_1^+;p_3^+;p_2^-) &= - S_R^\mu(p_2^-;p_3^+;p_1^+), \nonumber \\
S_R^\mu(p_1^-;p_3^-;p_2^+) &= \left[-S_R^\mu(p_2^-;p_3^+;p_1^+)\right]^*\,. 
\end{align}
It is also straightforward to include the spin-correlations with the leptonic decay products 
by contracting the partonic current with the leptonic current $L_\mu$ for fixed helicities of the
final state leptons. 
Consider the decay of the vector boson $V$ into two leptons:
$$V(q) \longrightarrow l^+(p_5) + l^-(p_6).$$
The purely vectorial tree-level leptonic current reads:
\begin{equation}
L^{\mu}(p_5,p_6) = \bar{u}(p_6) \, \gamma^{\mu} \, v(p_5),
\end{equation}
where in the case of an outgoing lepton-antilepton pair $L_\mu(p_5^-,p_6^+)$ 
corresponds to a right handed current, and $L_\mu(p_5^+,p_6^-)$ to a 
left-handed current.
We find straightforwardly:
\begin{equation}
L_{R}^\mu(p_5^-,p_6^+) =  [6 \;| \gamma^\mu |\; 5 \rangle, \qquad
L_{L}^\mu(p_5^+,p_6^-) =  [5 \;| \gamma^\mu |\; 6 \rangle = [L_{R}^\mu(p_5^-,p_6^+)]^*.
  \end{equation}

In order to write down the lepton-parton contraction
it is convenient to introduce the set of helicity coefficients defined in~\cite{3jtensor}:
\begin{align}
\alpha(u,v) &= \frac{s_{13} s_{23}}{4} \Big(2B + A_{12} - A_{11} \Big), \label{eqOmega1} \\
\beta(u,v)  &= \frac{s_{13} }{4} \Big(2 s_{23}B + 2(s_{12}+s_{13})A_{11} + s_{23}(A_{12}+A_{13} )  \Big), \label{eqOmega2} \\
\gamma(u,v) &= \frac{s_{13} s_{23}}{4} \Big( A_{11} - A_{13} \Big), \label{eqOmega3} \\
\delta(u,v) &= - \frac{s_{12} s_{13}}{4} A_{11}, \label{eqOmega4}
\end{align}
which, from their definition in terms of the coefficients $A_{ij}$ and $B$, 
respect the relation 
\begin{equation}
\alpha(u,v) - \beta(u,v) - \gamma(u,v) - \frac{2 s_{123}}{s_{12}} \delta(u,v) = 0.
\end{equation}
The relations above can be inverted for $A_{11}$, $A_{12}+2B$ and $A_{13}$, and
in these variables the contracted amplitude 
assumes a particularly simple form. 
We take the contraction of the right-handed quark current
with positive photon helicity, and the right-handed leptonic current, 
as basic object from which all other helicity configurations are obtained:
\begin{align}
&A_{RR}^{+}(p_5,p_6;p_1,p_3,p_2) = L_R^\mu(p_5^-;p_6^+) S_{R,\mu}(p_1^-;p_3^+;p_2^+) \notag \\
&= 
 - 2 \sqrt{2} \,
\Bigg[
\frac{  \langle 25 \rangle \langle 12 \rangle [16]}
            {\langle 13 \rangle \langle 23 \rangle}\, \alpha(u,v)
- \frac{\langle 25 \rangle [36]}
            {\langle 13 \rangle }\, \beta(u,v)
+\frac{\langle 15 \rangle [13]  [36]}{\langle 13 \rangle [23]}\, \gamma(u,v)\; \Bigg].
\end{align}

The unrenormalised helicity amplitude coefficients $\alpha$,
$\beta$ and $\gamma$ are vectors in colour space and
have perturbative expansions:
\begin{equation}
\Omega^{\rm un} = 
\sqrt{4 \pi \alpha} \, 
\delta_{ij} \, \left[
\Omega^{(0),{\rm un}}  
+ \left(\frac{\alpha_s}{2\pi}\right) \Omega^{(1),{\rm un}}  
+ \left(\frac{\alpha_s}{2\pi}\right)^2 \Omega^{(2),{\rm un}} 
+ {\cal O}(\alpha_s^3) \right] \;,
\end{equation}
for $\Omega = \alpha,\beta,\gamma$. The dependence on $(u,v)$ is again 
implicit.

From $A_{RR}^+(p_5,p_6;p_1,p_3,p_2)$, all other helicity amplitudes 
can be obtained by parity and charge conjugation. Axial contributions from the 
weak gauge boson couplings can be accounted for in a straightforward manner, 
by simply reweighting the different right-handed and left-handed
helicity amplitudes with appropriate weights. 

The eight possible helicity configurations are obtained from $A_{RR}^+$ as 
follows: 
\begin{align}
 L_R^\mu(p_5^-;p_6^+) S_{R\mu}(p_1^-;p_3^+;p_2^+) &= A_{RR}^+(p_5,p_6;p_1,p_3,p_2), \notag \nonumber \\ 
 L_R^\mu(p_5^-;p_6^+) S_{R\mu}(p_1^-;p_3^-;p_2^+) &= A_{RR}^-(p_5,p_6;p_1,p_3,p_2) =  [-A_{RR}^+(p_6,p_5;p_2,p_3,p_1) ]^*, \notag \nonumber \\
 L_R^\mu(p_5^-;p_6^+) S_{L\mu}(p_1^+;p_3^+;p_2^-) &= A_{RL}^+(p_5,p_6;p_1,p_3,p_2) =   -A_{RR}^+(p_5,p_6;p_2,p_3,p_1), \notag \nonumber \\
 L_R^\mu(p_5^-;p_6^+) S_{L\mu}(p_1^+;p_3^-;p_2^-) &= A_{RL}^-(p_5,p_6;p_1,p_3,p_2) =  [ A_{RR}^+(p_6,p_5;p_1,p_3,p_2) ]^*,  \\
\nonumber \\
 L_L^\mu(p_5^+;p_6^-) S_{R\mu}(p_1^-;p_3^+;p_2^+) &= A_{LR}^+(p_5,p_6;p_1,p_3,p_2) =    A_{RR}^+(p_6,p_5;p_1,p_3,p_2), \notag \nonumber \\
 L_L^\mu(p_5^+;p_6^-) S_{R\mu}(p_1^-;p_3^-;p_2^+) &= A_{LR}^-(p_5,p_6;p_1,p_3,p_2) =  [-A_{RR}^+(p_5,p_6;p_2,p_3,p_1) ]^*, \notag \nonumber \\
 L_L^\mu(p_5^+;p_6^-) S_{L\mu}(p_1^+;p_3^+;p_2^-) &= A_{LL}^+(p_5,p_6;p_1,p_3,p_2) =   -A_{RR}^+(p_6,p_5;p_2,p_3,p_1), \notag \nonumber \\
 L_L^\mu(p_5^+;p_6^-) S_{L\mu}(p_1^+;p_3^-;p_2^-) &= A_{LL}^-(p_5,p_6;p_1,p_3,p_2) =  [ A_{RR}^+(p_5,p_6;p_1,p_3,p_2) ]^*.
\end{align}

The general form of the gauge boson coupling to fermions is:
\begin{equation}
\mathcal{V}_\mu^{V,f_1f_2} = - i \,e\, \Gamma_{\mu}^{V,f_1f_2}  \qquad \mbox{with} \qquad e = \sqrt{4 \pi \alpha},
\end{equation}
whose explicit form depends on the gauge boson, on the type of 
fermions, and on their helicities:
\begin{equation}
\Gamma_{\mu}^{V,f_1f_2} = L_{f_1f_2}^V \; \gamma_\mu \left(1-\gamma_5 \over 2 \right) 
                     + R_{f_1f_2}^V \; \gamma_\mu \left(1+\gamma_5 \over 2 \right). 
\end{equation}
The left- and right-handed couplings are identical for a pure vector interaction, and 
are in general different if vector and axial-vector interactions contribute.
Their values for a photon are
\begin{equation}
L_{f_1f_2}^\gamma = R_{f_1f_2}^\gamma = -e_{f_1}\, \delta_{f_1f_2},
\end{equation}
while for a $Z$ boson
\begin{equation}
L_{f_1f_2}^Z = {I_3^{f_1} - \sin^2{\theta_w} e_{f_1} \over \sin{\theta_w} \cos{\theta_w}}\, \delta_{f_1f_2}\;,\qquad \qquad 
  R_{f_1f_2}^Z = -{\sin{\theta_w} e_{f_1} \over \cos{\theta_w}}\, \delta_{f_1f_2}\;,
  \end{equation}
and finally for a $W^\pm$
\begin{equation}
L_{f_1f_2}^{W} = { 1 \over \sqrt{2}\,\sin{\theta_w}}\;, \qquad \qquad R_{f_1f_2}^W = 0.
\end{equation}
The charges $e_i$ are measured in units of the fundamental electric charge $e>0$.

The vector boson propagator can be written as:
\begin{align}
P_{\mu \nu}^V(q,\xi) &= \frac{ i\, \Delta_{\mu \nu}^V(q,\xi) }{D_V(q)},
\end{align}
where  $\Delta_{\mu \nu}^{V}(q,\xi)$ and $D_V(q)$ are, respectively, the numerator 
and the denominator in the $R_\xi$ gauge:
\begin{align}
&\Delta_{\mu\nu}^{V}(q,\xi) = \left(-g_{\mu\nu} + (1-\xi) \frac{q_\mu q_\nu}{q^2-\xi M_V^2} \right),\label{numProp}
\end{align}
\begin{eqnarray}
 D_{Z,W^\pm}(q) &=& \big(\, q^2 - M_V^2 + i \Gamma_V M_V \, \big), \label{denProp} \\
 D_{\gamma}(q)  &=& q^2.  
\end{eqnarray}

In the narrow-width approximation we can simplify expression~\eqref{denProp} to 
\begin{equation}
D_{Z,W^\pm}(q) \approx i \Gamma_V M_V  \qquad \mbox{and} \qquad q^2 = M_V^2,
\end{equation}
where $M_V$ is the mass of the vector boson, while
$\Gamma_V$ is its decay width. 

Since we do not consider any electroweak corrections, the vector boson $V$
is always coupled to a fermion line and this allows us to neglect the $R_\xi$ dependence 
(or equivalently to put $\xi=1$).
With this notation we obtain for the different 
choices of $V = (\gamma^*,Z,W^\pm)$, and helicity combinations 
(with the obvious notation $p_{ij} = p_i + p_j$):
\begin{eqnarray}
\mathcal{M}_V(p_5^-,p_6^+;p_1^-,p_3^+,p_2^+) &=&
-i\,(4 \pi \alpha)\, \frac{\,R_{q_1q_2}^V\,R_{f_5f_6}^V\,}{D_V(p_{56})}\,   A_{RR}^+(p_5,p_6;p_1,p_3,p_2) ,\\
\mathcal{M}_V(p_5^-,p_6^+;p_1^+,p_3^-,p_2^-) &=&
-i\,(4 \pi \alpha)\, \frac{\,L_{q_1q_2}^V\,R_{f_5f_6}^V\,}{D_V(p_{56})}\, [ A_{RR}^+(p_6,p_5;p_1,p_3,p_2) ]^*,\\
\mathcal{M}_V(p_5^-,p_6^+;p_1^+,p_3^+,p_2^-) &=&
-i\,(4 \pi \alpha)\, \frac{\,L_{q_1q_2}^V\,R_{f_5f_6}^V\,}{D_V(p_{56})}\, [ - A_{RR}^+(p_5,p_6;p_2,p_3,p_1) ],\\
\mathcal{M}_V(p_5^-,p_6^+;p_1^-,p_3^-,p_2^+) &=&
-i\,(4 \pi \alpha)\, \frac{\,R_{q_1q_2}^V\,R_{f_5f_6}^V\,}{D_V(p_{56})}\, [ - A_{RR}^+(p_6,p_5;p_2,p_3,p_1) ]^*.
\end{eqnarray}
The corresponding amplitudes for left-handed leptonic current can be obtained simply
interchanging $p_5 \leftrightarrow p_6$ and $R_{f_5f_6}^V \rightarrow L_{f_5f_6}^V.$

\section{Outline of the calculation}
\label{sec:calc}
The two-loop corrections to the coefficients $\Omega$ 
for $W^\pm \gamma$ and $Z^0 \gamma$ production  
can be evaluated through a calculation of the relevant Feynman diagrams. 
The  diagrams which contribute at the two-loop level 
can be organised in different classes, 
some of which are present in both $V=Z^0,W^\pm$ cases, while others
contribute only to one of the two processes.

To identify the different classes of diagrams, it is useful to start with
the tree-level. We can define three classes of processes, each represented 
by a single diagram:
\begin{figure}[t]
\begin{center}
\includegraphics[height=3.0cm]{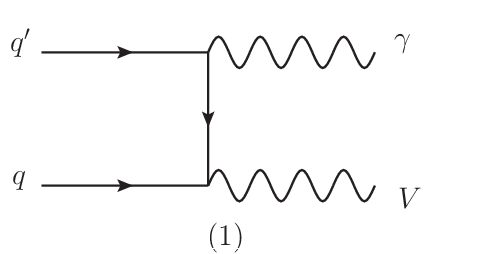}
\hspace{1cm}
\includegraphics[height=3.0cm]{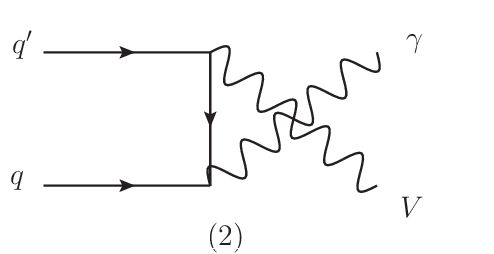}
\end{center}
\begin{center}
 \includegraphics[height=4.0cm]{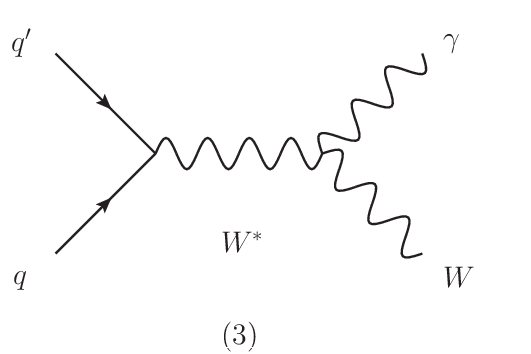}
\caption{Abelian and non-abelian tree-level contribution.}\label{treediag}
\end{center}
\end{figure}

\begin{enumerate}
 \item We call $I_1^{(0)}$ the contribution from diagram (1) 
 in Figure~\ref{treediag}, where the photon is attached on the 
 quark $q'$. The charge factor of this diagram is $e_{q'}$.
\item We refer as $I_2^{(0)}$ to the contribution from diagram (2) 
 in Figure~\ref{treediag},  where the photon is attached on
 the quark $q$. The charge factor of this diagram is  $e_{q}$.
\item $I_3^{(0)}$ is finally the contribution from diagram (3) 
 in Figure~\ref{treediag}, where an off-shell $W^*$ radiates
 the final state. The charge factor of this diagram is unity. 
\end{enumerate}
The $e_i$ are measured in units of $e$, so that $e_{q'} - e_{q} = 1$.

One can compute the $\Omega_j^{(0)}$ through the use of the projectors defined above, and
once the three different contributions are known one can reconstruct the correct values for the
helicity coefficients as:
\begin{equation}
\Omega_{W^\pm}^{(0)} = U_{qq'}\,\Big( e_{q'} \, \Omega_1^{(0)} + e_{q} \, \Omega_2^{(0)}
                                             +  \Omega_3^{(0)} \Big)
                     = U_{qq'}\,\Big[ e_{q'} \left(\Omega_1^{(0)} + \Omega_2^{(0)} \right) +  \Omega_3^{(0)} - \Omega_2^{(0)} \Big],
     \label{OmegaWgammaTree}
\end{equation}
\begin{equation}
\Omega_{Z,\gamma^*}^{(0)} = e_q \left( \Omega_1^{(0)} + \Omega_2^{(0)} \right) \label{OmegaZgammaTree},
\end{equation}
where $U_{ij}$ are the CKM matrix elements.
To simplify the expression above, we made use of the fact that the $Z^0$ boson does not couple to the photon,
and that in this case there is no flavour change, i.e.  $e_{q'} = e_q$.

At leading order, the $d$-dimensional expressions for the coefficients read:
\begin{align}
   \alpha_1^{(0)} &=
        \frac{\,d-4\,}{4(d-3)} \left(
            2
          + \frac{v}{u}
          - \frac{1}{u}
          \right), \qquad
     \alpha_2^{(0)} =
       - \frac{\,d-4\,}{4(d-3)} \left(
            2
          + \frac{v}{u}
          - \frac{1}{u}
          \right)
       + 1,\nonumber\\
   \alpha_3^{(0)} &=
       - \frac{\,d-4\,}{4(d-3)} \left(
            2
          + \frac{v}{u}
          - \frac{1}{u}
          \right)
       + 1 - \frac{u}{1-v}.
\end{align}
\begin{align}
   \beta_1^{(0)} &=
        \frac{\,d-4\,}{4(d-3)} \left(
            1
          - \frac{1}{u}
          \right)
       + 1, \qquad
    \beta_2^{(0)} =
       - \frac{\,d-4\,}{4(d-3)} \left(
            1
          - \frac{1}{u}
          \right),\nonumber \\
   \beta_3^{(0)} &=
       - \frac{\,d-4\,}{4(d-3)} \left(
            1
          - \frac{1}{u}
          \right)
       - \frac{u}{1-v}.
\end{align}
\begin{align}
   \gamma_1^{(0)} &=
       - \frac{\,d-4\,}{4(d-3)} \left(
            1
          + \frac{v}{u}
          \right), \qquad
    \gamma_2^{(0)} =
        \frac{\,d-4\,}{4(d-3)} \left(
            1
          + \frac{v}{u}
          \right),\nonumber \\
   \gamma_3^{(0)} &=
        \frac{\,d-4\,}{4(d-3)} \left(
            1
          + \frac{v}{u}
          \right).
\end{align}

Using~\eqref{OmegaWgammaTree} and \eqref{OmegaZgammaTree} we find
\begin{align}
 \alpha_W^{(0)} &= U_{qq'}\left( e_{q'} - \frac{u}{1-v} \right), 
 \qquad \beta_W^{(0)} = U_{qq'}\left(e_{q'} - \frac{u}{1-v}\right), 
 \qquad \gamma_W^{(0)} = 0,
\end{align}
\begin{align}
 \alpha_{Z,\gamma^*}^{(0)} = e_q, \qquad \beta_{Z,\gamma^*}^{(0)} = e_q, \qquad \gamma_{Z,\gamma^*}^{(0)} = 0.
\end{align}

\begin{figure}[t]
\begin{center}
\includegraphics[height=3.0cm]{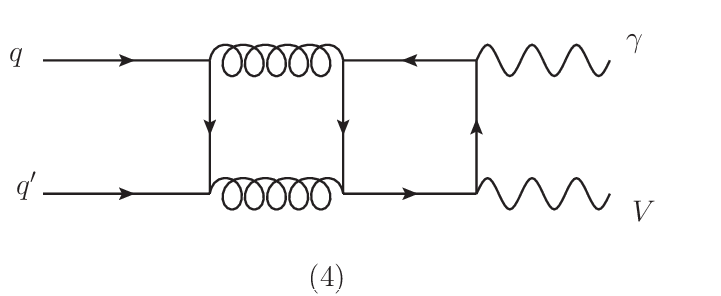}
\end{center}
\begin{center}
\includegraphics[height=4.0cm]{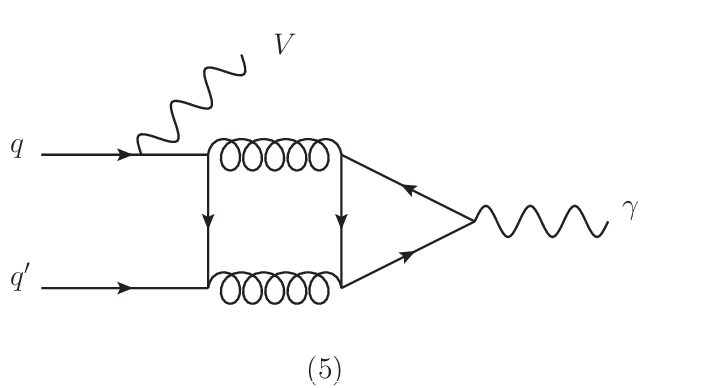}
\includegraphics[height=4.0cm]{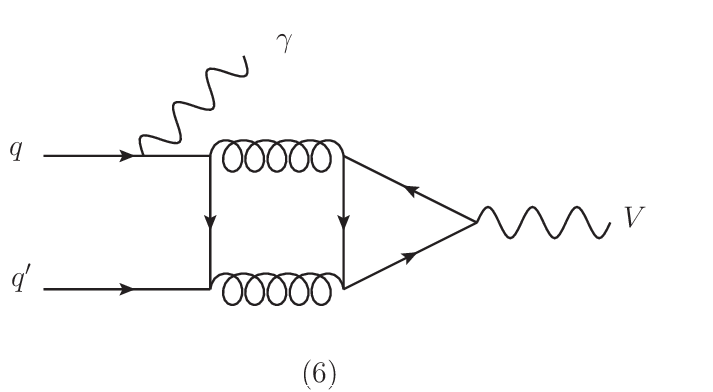}
\caption{Examples of two-loop diagrams in the classes $I_4^{(2)}$, $I_5^{(2)}$ and $I_6^{(2)}$.}\label{2ldiag}
\end{center}
\end{figure}

At one-loop, the same classification of contributions applies:
$I_1^{(1)},I_2^{(1)},I_3^{(1)}$. A further type of diagrams, 
with both the photon and the gauge boson 
which couple to a closed quark loop, is zero due to colour conservation. 
At two loops, besides $I_1^{(2)},I_2^{(2)},I_3^{(2)},$ three further classes of diagrams appear:
\begin{enumerate}
 \item $I_4^{(2)}$ are the diagrams where both $\gamma$ and $V$ 
       couple to the same fermion loop, as depicted for example in Figure~\ref{2ldiag}, diagram (4). 

       This contribution is denoted
       by $\NFZ$ and is proportional
       to the charge weighted  sum of the quark flavours. 
       In case of a $\gamma^*$ exchange we find
       \begin{equation}
        N_{F,\gamma} = \frac{\sum_{q'} \, e^2_{q'}}{e_q}\;.
       \end{equation}

       Considering $Z$-interactions, the same class of diagrams yields not only a 
       contribution from the  vector component of the $Z$, which for the right-handed
       quark amplitude is given by
       \begin{equation}
        N_{F,Z} = \frac{\sum_{q'} \left(L^Z_{q'q'}+R^Z_{q'q'}\right)e_{q'}}{2R^Z_{qq}}\; ,
       \end{equation}
       but also a contribution involving the axial couplings of the $Z$. 
       This contribution vanishes identically for $Z^0\gamma$ production,
       already before summing over the quark flavours inside the loop.
       In the case of $W^\pm$ exchange charge conservation ensures that
       \begin{equation}
        N_{F,W^\pm} = 0.
       \end{equation}
 \item $I_5^{(2)}$ are the diagrams where the photon couples alone to a fermion loop, while
       $V$ couples to the fermion line, as depicted in Figure~\ref{2ldiag}, diagram (5). 
       This class of diagrams has to sum to zero 
       due to Furry's theorem.
 \item $I_6^{(2)}$ are finally the diagrams where $V$ couples alone to a fermion loop, while
       $\gamma$ couples to the fermion line, as depicted in Figure~\ref{2ldiag}, diagram (6). 
       These diagrams give both a vector and an axial contribution, where the vector contribution
       is again zero due to Furry's theorem,
       while the axial contribution is zero in the framework of massless QCD~\cite{3jtensor}.
  \end{enumerate}     
  
       We explicitly evaluated the contributions from classes $I_5^{(2)},I_6^{(2)}$, 
       and their vanishing provides a check on our calculation.

The classes $I_j^{(2)},\; j=1,6\,$ exhaust all the possible two-loop QCD diagrams which
can contribute to the production of a pair $V\, \gamma$, 
whatever is the identity of the vector boson $V$.\newline
When computing the helicity amplitudes, one can  evaluate the contributions from these six
classes of diagrams independently, without keeping track
of the axial current contributions, thus considering the vector boson $V$ as an off-shell 
purely vector particle. 
Once the $I_j^{(2)}$ are known, as for the tree-level case,
one can reconstruct the proper amplitudes for $V= Z^0, W^\pm$
summing these six contributions up, 
multiplied by appropriate weights.\newline

The calculation proceeds as follows.
The $143$ two-loop diagrams belonging to the classes $j=(1,2,4,5,6)$ are produced 
using QGRAF~\cite{qgraf} while, on the other hand, 
we did not need to evaluate explicitly the
diagrams in class $I_3^{(2)}$, since they only account for the QCD
corrections of the quark form factor, which is known up to three loops
in the literature~\cite{formfactor1,formfactor2}.

The tensor coefficients are then evaluated analytically diagram by diagram 
applying the projectors defined in~\cite{3jtensor}. 
As a result, one obtains the tensor coefficients in terms of thousands of 
planar and non-planar two-loop scalar integrals, which can be easily
classified in two auxiliary topologies, one planar and the other non-planar~\cite{3jme}.
Through the usual IBP identities~\cite{chet1} one can reduce independently all the integrals
belonging to these two auxiliary topologies to a small set
of master integrals. This reduction is performed using the Laporta algorithm~\cite{laporta}, 
implemented in the Reduze code~\cite{reduze}.
All the masters for such topologies are known~\cite{3jmaster} as series in
the parameter $\epsilon = (4-d)/2$, through a systematic approach based
on the differential equation method~\cite{de2,3jmaster}. The masters are 
expressed as Laurent expansion in $\e$, with coefficients containing 
harmonic polylogarithms (HPLs,~\cite{hpl}) and two-dimensional harmonic 
polylogarithms (2dHPLs,~\cite{3jmaster}). Numerical implementations of these 
functions are available~\cite{hplnum}. 
For all the intermediate algebraic manipulations we have made extensive use of 
FORM~\cite{form}.

The two-loop unrenormalised helicity coefficients $\Omega^{(2),{\rm un}}$
can then be evaluated as linear combination
of the tensor coefficients, and in particular they can be evaluated separately 
for every class of diagrams:
\begin{equation}
  \Omega_j^{(2),{\rm un}}\;, \qquad \mbox{with}\; j=1,6\,.
\end{equation}
As for the tree level case, 
it is trivial to reconstruct the amplitudes for the processes considered as linear combinations
of these six amplitudes.

We start considering the case where $V = W^\pm$. The $W$ boson couples only to
left-handed fermions and charge conservation implies that $N_{F,W} = 0$.
The amplitudes for $W^\pm \gamma$ production at two loops thus
receive contributions only from three of the six classes of diagrams above, i.e.:
\begin{align}
 \Omega_{W^\pm}^{(2),{\rm un}} &= U_{q q'}\,\Big(\, e_{q'}\, \Omega_1^{(2),{\rm un}} + e_q \, \Omega_2^{(2),{\rm un}} 
                                                   +  \Omega_3^{(2),{\rm un}} \, \Big)\nonumber \\
                               &= U_{q q'}\,\Big[\,e_{q'}\, \left(\Omega_1^{(2),{\rm un}} +  \Omega_2^{(2),{\rm un}} \right)
                                                   +  \Omega_3^{(2),{\rm un}} - \Omega_2^{(2),{\rm un}} \Big].
\end{align}
As for the tree level, $U_{ij}$ are the CKM matrix elements.\newline

For the $V = Z^0,\gamma^*$ case, $I_3^{(n)}=0$ at all orders, since the $Z^0$ and $\gamma^*$ 
are electrically neutral, while  $I_4^{(2)}$ is non-vanishing
since charge conservation does not forbid such diagrams anymore. 
\begin{equation}
 \Omega_{V}^{(2),{\rm un}} = e_q\, \left( \Omega_1^{(2),{\rm un}} 
                         +\,\Omega_2^{(2),{\rm un}} \right)+\,N_{F,V}\,\Omega_4^{(2),{\rm un}},
\end{equation}
for $V=(Z^0,\gamma^*)$.

\section{Two-loop helicity amplitudes}
\label{sec:results}
Renormalisation of ultraviolet divergences is 
performed in the \MSbar\ scheme by replacing 
the bare coupling $\alpha_0$ with the renormalised coupling 
$\alpha_s\equiv \alpha_s(\mu^2)$,
evaluated at the renormalisation scale $\mu^2$. 
Since the tree amplitudes are of 
${\cal O}(\alpha_s^0)$,
we only need the one loop relation between the bare and renormalised couplings:
\begin{equation}
\alpha_0\mu_0^{2\e} S_\e = \alpha_s \mu^{2\e}\left[
1- \frac{\beta_0}{\e}\left(\frac{\alpha_s}{2\pi}\right) 
+{\cal O}(\alpha_s^2) \right]\; ,
\end{equation}
where
\begin{displaymath}
S_\e =(4\pi)^\e e^{-\e\gamma}\qquad \mbox{with Euler constant }
\gamma = 0.5772\ldots
\end{displaymath}
and $\mu_0^2$ is the mass parameter introduced 
in dimensional regularisation to maintain a 
dimensionless coupling 
in the bare QCD Lagrangian density.
$\beta_0$ is the first 
coefficient of the QCD $\beta$-function:
\begin{equation}
\beta_0 = \frac{11 \CA - 4 T_R \NF}{6},
\end{equation}
with the QCD colour factors
\begin{equation}
\CA = N,\qquad C_F = \frac{N^2-1}{2N},
\qquad T_R = \frac{1}{2}\; .
\end{equation}
The renormalisation is performed at fixed scale $\mu^2 = q^2$. The renormalised 
helicity coefficients read:
\begin{eqnarray}
\Omega^{(0)}  &=& \Omega^{(0),{\rm un}} ,
 \nonumber \\
\Omega^{(1)}  &=& 
S_\e^{-1} \Omega^{(1),{\rm un}},  \nonumber \\
\Omega^{(2)} &=& 
S_\e^{-2} \Omega^{(2),{\rm un}}  
-\frac{\beta_0}{\e} S_\e^{-1}
\Omega^{(1),{\rm un}}\;.
\end{eqnarray}
The full scale dependence 
of the coefficients can be recovered from the renormalisation group. It reads:
\begin{eqnarray}
\Omega (\mu^2,\alpha_s(\mu^2)) &= &
\sqrt{4 \pi \alpha} \, 
\delta_{ij} \, \Bigg[
\Omega^{(0)}
+ \left(\frac{\alpha_s(\mu^2)}{2\pi}\right) \Omega^{(1)}  \nonumber \\
&& + \left(\frac{\alpha_s(\mu^2)}{2\pi}\right)^2 \left(\Omega^{(2)} 
+ \beta_0 \Omega^{(1)} \ln \left(\frac{\mu^2}{q^2}\right)\right)
+ {\cal O}(\alpha_s^3) \Bigg] \;,
\end{eqnarray}

After performing ultraviolet renormalisation,
the amplitudes still
contain singularities, which are of infrared origin and will be  analytically
cancelled by those occurring in radiative processes of the
same order.
Catani~\cite{catani} has shown how to organise the 
infrared pole structure of the one- and two-loop contributions renormalised in the 
$\overline{{\rm MS}}$-scheme in terms of the tree and renormalised one-loop amplitudes.
The same procedure applies to the tensor coefficients. Their pole structure can be 
separated off as follows:
\begin{eqnarray}
\Omega^{(1)} &=& {\bom I}^{(1)}(\epsilon) \Omega^{(0)} +
\Omega^{(1),{\rm finite}},\nonumber \\
\Omega^{(2)} &=& \Biggl (-\frac{1}{2}  {\bom I}^{(1)}(\epsilon) {\bom I}^{(1)}(\epsilon)
-\frac{\beta_0}{\epsilon} {\bom I}^{(1)}(\epsilon) 
+e^{-\epsilon \gamma } \frac{ \Gamma(1-2\epsilon)}{\Gamma(1-\epsilon)} 
\left(\frac{\beta_0}{\epsilon} + K\right)
{\bom I}^{(1)}(2\epsilon) + {\bom H}^{(2)}(\epsilon) 
\Biggr )\Omega^{(0)}\nonumber \\
&& + {\bom I}^{(1)}(\epsilon) \Omega^{(1)}+ \Omega^{(2),{\rm finite}},
\end{eqnarray}
where the constant $K$ is
\begin{equation}
K = \left( \frac{67}{18} - \frac{\pi^2}{6} \right) \CA - 
\frac{10}{9} T_R \NF.
\end{equation}

In our case, there is only a 
quark--antiquark pair present in the initial
state, so that
$\bom{I}^{(1)}(\epsilon)$ 
is given by,
\begin{equation}
\bom{I}^{(1)}(\epsilon)
=
- \frac{e^{\epsilon\gamma}}{2\Gamma(1-\epsilon)} \Biggl[
\frac{N^2-1}{2N}
\left(\frac{2}{\epsilon^2}+\frac{3}{\epsilon} \right)
{\tt S}_{12}\Biggr ]\; ,\label{eq:I1}
\end{equation}
where, since we have set $\mu^2 = s_{123}^2$:
\begin{equation}
{\tt S}_{12} = \left(-\frac{\mu^2}{s_{12}}\right)^{\epsilon}
= \left( x \right)^{-\epsilon} \left(-1 - i 0 \right)^{-\epsilon}.
\end{equation}
Note that on expanding ${\tt S}_{12}$,
imaginary parts are generated, 
the sign of which is fixed by the small imaginary
part $+i0$ of $\sab$.
The hard radiation constant is a scalar in colour space: 
\begin{equation}
{\bom H}^{(2)}(\epsilon) = \frac{e^{\epsilon \gamma}}{4\,\epsilon\,\Gamma(1-\epsilon)} H^{(2)}.
\end{equation}
with
\begin{equation}
H^{(2)} =  2H^{(2)}_{q}
\end{equation}
where in the \MSbar\ scheme
\begin{eqnarray}
H^{(2)}_q &=& (N^2-1)\, \Bigg[
\left({7\over 4}\zeta_3+{\frac {409}{864}}- {\frac {11\pi^2}{96}}\right)
+\left({3\over 2}\zeta_3+{3\over 32}-{\pi^2\over 8}\right){1\over N^2}\nonumber \\
&&+\left({\pi^2\over 48}-{25\over 216}\right){N_F\over N}\Bigg]\;.
\end{eqnarray}

For the infrared factorisation of the two-loop results, 
the renormalised next-to-leading order helicity amplitude coefficients
are needed through to ${\cal O}(\epsilon^2)$.
Their decomposition in colour structures is straightforward:
\begin{equation}
\Omega^{(1),{\rm finite}}_{j}(u,v) =  
C_F\, a_{\Omega}^{(j)}(u,v)
   \;. 
\label{eq:oneloopamp}
\end{equation}
The expansion of the coefficients through to $\e^2$ yields HPLs and 2dHPLs 
up to weight 4. The explicit expressions are of considerable size, 
such that we only quote the $\e^0$-terms in the appendix. To this order, 
the coefficients had been derived previously~\cite{ert1,gg} in terms of 
logarithms and dilogarithms. 
The expressions through to ${\cal O}(\e^2)$
in FORM format are appended to the arXiv submission 
of this article.

The finite two-loop remainder is obtained by subtracting the
predicted infrared structure (expanded through to ${\cal O}(\epsilon^0)$) from
the renormalised helicity coefficient.  We further decompose the 
finite remainder according to the colour structures as follows: 
\begin{align}
\Omega^{(2),{\rm finite}}_j(u,v) &= 
   (N^2 - 1)\,\left( A_\Omega^{(j)}(u,v) + \frac{1}{N^2} B_\Omega^{(j)}(u,v) \right) 
 + C_F\, \NF C_\Omega^{(j)}(u,v) \nonumber \\ 
&+ C_F\, \NFZ D_\Omega^{(j)}(u,v) \; , 
\label{eq:twoloopamp}
\end{align}
where the last term, as discussed above, 
is generated by graphs where the virtual gauge boson does not
couple directly to the final-state quarks, and is different from 
zero only for the $V=Z,\gamma^*$ case.   

The helicity coefficients contain HPLs and 2dHPLs up to weight 4.
The size of each helicity coefficient is comparable 
to the size of the helicity-averaged tree times two-loop matrix element 
for $3j$ production quoted in~\cite{3jme}, and we decided not to
include them here explicitly. The complete set of coefficients in FORM 
format is attached to the arXiv submission of this article.

\section{Checks on the result}
\label{sec:checks}
Several non-trivial checks were applied to validate our results:
\begin{enumerate}
 \item All seven tensor coefficients in  (\ref{eq:smubase}) were computed. We validated 
       that they fulfil the following relations, which follow 
       from the symmetry properties of the tensor under an interchange
       $p_1 \leftrightarrow p_2$:
       \begin{eqnarray}
        A_{21}(\sac,\sbc,\sabc) &=& - A_{12}(\sbc,\sac,\sabc),\nonumber \\
        A_{22}(\sac,\sbc,\sabc) &=& - A_{11}(\sbc,\sac,\sabc),\nonumber \\
        A_{23}(\sac,\sbc,\sabc) &=& - A_{13}(\sbc,\sac,\sabc),\nonumber \\
        B(\sac,\sbc,\sabc) &=& B(\sbc,\sac,\sabc).
       \end{eqnarray}
 \item We computed explicitly the helicity coefficients for all the diagrams 
       in classes $I^{(2)}_5$ and $I^{(2)}_6$, which should yield a vanishing contribution due to 
       Furry's theorem. Each diagram gives a non-vanishing  contribution, a full cancellation is 
       obtained only in the sum of all diagrams.
 \item The IR singularity structure of our result agrees with the prediction 
       of the Catani formula~\cite{catani}.
 \item We compared the helicity coefficients 
       $\Omega_Z^{(2)}$ for $q \bar{q} \rightarrow Z \gamma$,
       with those for $\gamma^* \rightarrow q \bar{q} g$~\cite{3jtensor}. As explained in 
       Appendix~\ref{app:vjet}, 
       the unrenormalised two-loop $\Omega_{3j}$ 
       coefficients can be decomposed according to their colour structure as:
      \begin{align}
      \Omega_{3j}^{(2,{\rm un})}(u,v) &=  
      N^2 A^{(3j,{\rm un})}_\Omega(u,v) + B^{(3j,{\rm un})}_\Omega(u,v) + \frac{1}{N^2} C^{(3j,{\rm un})}_\Omega(u,v) \nonumber \\ 
      &+ N\NF D^{(3j,{\rm un})}_\Omega(u,v) 
      + \frac{\NF}{N} E^{(3j,{\rm un})}_\Omega(u,v) + \NF^2 F^{(3j,{\rm un})}_\Omega (u,v)
      \nonumber \\ & +   \NFZ \left(\frac{4}{N}-N\right) G^{(3j,{\rm un})}_\Omega(u,v) \;.
      \end{align}
      We checked that in the decay kinematical configuration $Z \rightarrow q \, \bar{q} \, \gamma$, 
        prior to  UV renormalisation and
        IR subtraction, the following identities are fulfilled
       \begin{equation}
         \begin{array}{cccc}  B_\Omega^{(Z,{\rm un})}(y,z) &=& -& C^{(3j,{\rm un})}_\Omega(y,z)\;,\\
                                  C_\Omega^{(Z,{\rm un})}(y,z) &=& -& 2\,E^{(3j,{\rm un})}_\Omega(y,z)\;, \\
                                  D_\Omega^{(Z,{\rm un})}(y,z) &=& -& 4\,G^{(3j,{\rm un})}_\Omega(y,z)\;, \end{array} 
       \end{equation}
       which follow from the structure of the underlying two-loop diagrams. UV renormalisation 
       and IR subtraction of the  $Z\to q\bar q \gamma$ and $Z\to q \bar q g$ amplitudes 
       differ. However, two of the above relations are unaffected by renormalisation and 
       retain the same IR structure, such that we obtain: 
       \begin{equation}
        \begin{array}{cccc} B_\Omega^{(Z,{\rm finite})}(u,v) &=& -& C^{(3j,{\rm finite})}_\Omega(u,v)\;,  \\
        D_\Omega^{(Z,{\rm finite})}(u,v) &=& -& 4\,G^{(3j,{\rm finite})}_\Omega(u,v)\;,\end{array} 
       \end{equation}
       which also remain true after analytic continuation. 
\end{enumerate}

\section{Conclusions and Outlook}
\label{sec:conc}
In this paper we derived the two-loop corrections to the helicity amplitudes 
for the processes $q\bar q \to W^\pm \gamma$ and 
$q\bar q \to Z^0\gamma$. Our calculation was 
performed in dimensional regularisation by applying $d$-dimensional 
projection operators to the most general tensor structure of the amplitude. 
Our results are expressed in terms of dimensionless helicity coefficients, which multiply 
the basic tree-level amplitudes, expressed in four-dimensional spinors. 
By applying Catani's infrared factorisation 
formula, we extract the finite parts of the helicity coefficients, which 
are independent on the precise scheme used to define the helicity amplitudes. 
We provide compact analytic expressions for the two-loop helicity coefficients 
in terms of HPLs and 2dHPLs. 

These amplitudes are relevant to the NNLO corrections to $V\gamma$ production 
at hadron colliders. This process provides a direct access to the 
photonic couplings of the weak gauge bosons and is a crucial test of the 
structure of the electroweak theory at high energies. With the 
amplitudes derived here, in combination with the amplitudes relevant to 
$V\gamma j$ production at NLO~\cite{vgjnlo}, all ingredients to this 
NNLO calculation are now
available. Since the leading order contribution to this process does not 
contain any QCD partons in the final state, the 
well-established $q_T$ subtraction method~\cite{qtsub} 
could be used for this calculation.

\section*{Acknowledgements}
This research was supported in part by
the Swiss National Science Foundation (SNF) under contract
PDFMP2-135101 and  by the European Commission through the 
``LHCPhenoNet" Initial Training Network PITN-GA-2010-264564.  

\appendix
\section{One-loop helicity amplitudes}
We list here the $\e^0$-terms of the one-loop helicity coefficients for the groups
of diagrams defined above:

{\small
\begin{align}
   a_{\alpha}^{(1)} &=
        \zeta_2
          + G( - v,u)i\,\pi\,
          - G( - v,u)H(0,v)
          + G( - v,0,u)
          + G(0,u)H(0,v)
          + H(0,v)i\,\pi \nonumber \\
        &- H(0,0,v)
          + H(1,0,v) + {\cal O} (\e)\; ,\nonumber \\ 
   a_{\alpha}^{(2)} &=
         \frac{(1-v)^2}{u^2}\,\Bigg[
            G(1 - v,u)H(0,v)
          + G(1,1 - v,u)
          + G(1,u)i\,\pi\,
          - G(1,u)H(0,v) \nonumber  \\
          &- G(1,u)H(1,v)
          \Bigg] 
        +\, \frac{(1-u-v)( 2 - 2v - u + 3uv - u^2)}{2u(1-u)^2}\, \,\Bigg[
            i\,\pi\,
          + G(1 - v,u)
          - H(1,v)
          \Bigg] \nonumber \\
          &+\, \frac{v(2 - 2v - 2u + 3uv)}{2u(1-u)^2}\, H(0,v) 
          +\frac{(7u - 7 - v)}{2(1-u)} + {\cal O} (\e)\; , \nonumber \\ 
   a_{\alpha}^{(3)} &=
         - \frac{4(1-u-v)}{(1-v)} + {\cal O} (\e)\; ,\\  \nonumber \\
   a_{\beta}^{(1)} &=
          \zeta_2
          + G( - v,u)i\,\pi\,
          - G( - v,u)H(0,v)
          + G( - v,0,u)
          + G(0,u)H(0,v)
          + H(0,v)i\,\pi\,  \nonumber \\
        & - H(0,0,v)
          + H(1,0,v)
        + \frac{v( - 3 + 3v + u)}{2(1-v)^2}\,
          H(0,v)\,
        +\, \frac{( - 9 + 9v + u)}{2(1-v)} + {\cal O} (\e)\; \nonumber  \\ 
   a_{\beta}^{(2)} &=
         \frac{(1 - v + uv)}{u^2}\,\,\Bigg[
            G(1 - v,u)H(0,v)
          + G(1,1 - v,u)
          + G(1,u)i\,\pi\,
          - G(1,u)H(0,v)  \nonumber \\
         &- G(1,u)H(1,v)
          \Bigg]
           +  \frac{(2 - 2v - u + 3uv - u^2)}{2u(1-u)}\, \,\Bigg[
            i\,\pi\,
          + G(1 - v,u)
          - H(1,v)
          \Bigg]  \nonumber \\
        &+ \frac{v(2-4v+2v^2+3uv-3uv^2-2u^2+3u^2v-u^3) }{2u(1-u)(1-v)^2}\,
          H(0,v)
         +\,  \frac{(1 - v + u)}{2(1-v)}\,+ {\cal O} (\e)\; ,
       \nonumber    \\
   a_{\beta}^{(3)} &=
         \frac{4u}{(1-v)} + {\cal O} (\e)\; , \\  \nonumber \\ 
   a_{\gamma}^{(1)} &=
         \frac{v(1-u-v)}{2(1-v)^2} \, H(0,v)
         + \frac{(1-u-v)}{2(1-v)} + {\cal O} (\e)\; , \nonumber \\ 
   a_{\gamma}^{(2)} &=
      \frac{v(1-u-v)}{u^2}\,\Bigg[
            G(1 - v,u)H(0,v)
          + G(1,1 - v,u)
          + G(1,u)i\,\pi\,
          - G(1,u)H(0,v) \nonumber  \\
         &- G(1,u)H(1,v)
          \Bigg]
         + \frac{v(1-u-v)( 2 - 3u)}{2u(1-u)^2}\, \,\Bigg[
            i\,\pi\,
          + G(1 - v,u)
          - H(1,v)
          \Bigg]\, \nonumber \\ &+ \frac{(1-u-v)( v - u)}{2(1-u)(1-v)}
         + \frac{v(1-u-v)(2v-2v^2-2uv+3uv^2-2u^2v+u^3) }{2u(1-u)^2(1-v)^2}\,H(0,v)
         + {\cal O} (\e)\; , \nonumber \\
   a_{\gamma}^{(3)} &= {\cal O} (\e)\; .
\end{align}
}

It should also be noted that these finite pieces of the one-loop coefficients
can equally be written in terms of ordinary logarithms and dilogarithms, 
see~\cite{ert1,gg}.
The reason to express them in terms of HPLs and 2dHPLs 
here is their usage in the infrared counter-term of the two-loop coefficients, 
which cannot be fully expressed in terms of logarithmic and polylogarithmic 
functions.

\section{Two-loop leading colour amplitudes}
The analytical expressions for the 
finite remainders of the two-loop amplitudes, as defined in 
(\ref{eq:twoloopamp}), are of considerable size. For illustration, 
we only quote the leading colour contributions here.

{\small
\begin{align*}
   A_\alpha^{(1)} &=
          \frac{1}{4} \, \Big[
          - G( - v,1 - v, - v,u)\,i\,\pi
          + G( - v,1 - v, - v,u)\,H(0,v)
          - G( - v,1 - v, - v,0,u) \\ &
          - G( - v,1 - v,u)\,\zeta_2
          - G( - v,1 - v,u)\,H(0,v)\,i\,\pi
          + G( - v,1 - v,u)\,H(0,0,v) \\ &
          - G( - v,1 - v,u)\,H(1,0,v)
          - G( - v,1 - v,0,u)\,H(0,v)
          - 2\,G( - v, - v, - v,u)\,i\,\pi \\ &
          + 2\,G( - v, - v, - v,u)\,H(0,v)
          - 2\,G( - v, - v, - v,0,u)
          - 4\,G( - v, - v,u)\,\zeta_2 \\ &
          - 2\,G( - v, - v,u)\,H(0,v)\,i\,\pi
          + 2\,G( - v, - v,u)\,H(0,0,v)
          + 2\,G( - v, - v,0,u)\,i\,\pi \\ &
          - 2\,G( - v, - v,0,u)\,H(0,v)
          + 2\,G( - v, - v,0,0,u)
          - 2\,G( - v,u)\,\zeta_3 \\ &
          - G( - v,u)\,i\,\pi\,\zeta_2
          - 3\,G( - v,u)\,H(0,0,v)\,i\,\pi
          + 3\,G( - v,u)\,H(0,0,0,v) \\ &
          - G( - v,u)\,H(0,1,0,v)
          + G( - v,u)\,H(1,v)\,\zeta_2
          + G( - v,u)\,H(1,0,v)\,i\,\pi \\ &
          - G( - v,u)\,H(1,0,0,v)
          + G( - v,u)\,H(1,1,0,v)
          + 3\,G( - v,0, - v,u)\,i\,\pi \\ &
          - 3\,G( - v,0, - v,u)\,H(0,v)
          + 3\,G( - v,0, - v,0,u)
          - G( - v,0,u)\,\zeta_2 \\ &
          + 2\,G( - v,0,u)\,H(0,v)\,i\,\pi
          - 3\,G( - v,0,u)\,H(0,0,v)
          + G( - v,0,u)\,H(1,0,v) \\ &
          + 2\,G( - v,0,0,u)\,H(0,v)
          + 2\,G(0,1 - v, - v,u)\,i\,\pi
          - 2\,G(0,1 - v, - v,u)\,H(0,v) \\ &
          + 2\,G(0,1 - v, - v,0,u)
          + 2\,G(0,1 - v,u)\,\zeta_2
          + 2\,G(0,1 - v,u)\,H(0,v)\,i\,\pi \\ &
          - 2\,G(0,1 - v,u)\,H(0,0,v)
          + 2\,G(0,1 - v,u)\,H(1,0,v)
          + 2\,G(0,1 - v,0,u)\,H(0,v) \\ &
          + 2\,G(0, - v, - v,u)\,i\,\pi
          - 2\,G(0, - v, - v,u)\,H(0,v)
          + 2\,G(0, - v, - v,0,u) \\ &
          - 2\,G(0, - v,u)\,\zeta_2
          + G(0, - v,u)\,H(0,v)\,i\,\pi
          - 2\,G(0, - v,u)\,H(0,0,v) \\ &
          + G(0, - v,0,u)\,H(0,v)
          - G(0,u)\,H(0,v)\,\zeta_2
          + 2\,G(0,u)\,H(0,0,v)\,i\,\pi \\ &
          - 3\,G(0,u)\,H(0,0,0,v)
          + G(0,u)\,H(0,1,0,v)
          + 2\,G(0,u)\,H(1,0,0,v)
          - 2\,G(0,0, - v,u)\,i\,\pi \\ &
          + 2\,G(0,0, - v,u)\,H(0,v)
          - 2\,G(0,0, - v,0,u) 
          + 2\,G(0,0,u)\,H(0,0,v)
          - 2\,H(0,v)\,\zeta_3 \\ &
          - H(0,v)\,i\,\pi\,\zeta_2 
          - 3\,H(0,0,0,v)\,i\,\pi 
          + 3\,H(0,0,0,0,v)
          - H(0,0,1,0,v) 
          + H(0,1,v)\,\zeta_2  \\ &
          + H(0,1,0,v)\,i\,\pi 
          - H(0,1,0,0,v) 
          + H(0,1,1,0,v)
          + 2\,H(1,v)\,\zeta_3  \\ &
          + 2\,H(1,0,v)\,\zeta_2 
          + 2\,H(1,0,0,v)\,i\,\pi
          - 2\,H(1,0,0,0,v)
          + 2\,H(1,0,1,0,v)
          + 2\,H(1,1,0,0,v)
          \Big]\\
       &- \frac{5}{16} \,
           \zeta_4
        + \frac{11}{12} \, \Big[
          - G( - v, - v,u)\,i\,\pi
          + G( - v, - v,u)\,H(0,v)
          - G( - v, - v,0,u)\\ &
          - G( - v,u)\,H(0,0,v)
          + G( - v,0,u)\,H(0,v)
          - G( - v,0,0,u)
          - G(0,0,u)\,H(0,v)
          \Big]\\
       &- \frac{5}{4} \, \Big[
            G( - v,u)\,H(0,v)\,\zeta_2
          + H(0,0,v)\,\zeta_2
          \Big]
        - \frac{11}{6} \, 
          G( - v,u)\,\zeta_2 \\
       &+ \frac{v}{16\,(u+v)^2}\,\Big(1 - 3\,v - 3\,u\, \Big) \, 
           H(0,v)
          \\
       &+ \frac{v}{8\,u\,(1-u)^3}\,\Big(2 - 2\,v - 4\,u + 5\,u\,v + 4\,u^2 - 5\,u^2\,v
       - 2\,u^3\, \Big) \,  \Big[
          - G(0, - v,u)\,i\,\pi\\ &
          + G(0, - v,u)\,H(0,v)
          - G(0, - v,0,u)
          \Big]\\
       &+ \frac{(1-u-v)}{8\,u\,(1-u)^3}\,\Big(2 - 2\,v - 3\,u + 5\,u\,v - 5\,u^2\,v
       + u^3\, \Big) \,  \Big[
          - 2\,i\,\pi\,\zeta_2
          - H(1,0,v)\,i\,\pi
          \Big]\\
       &+ \frac{(1-u-v)}{4\,u^2}\,\Big(1 - v + u\, \Big) \, \Big[
          + G(1 - v,1 - v, - v,u)\,i\,\pi
          - G(1 - v,1 - v, - v,u)\,H(0,v) \\ &
          + G(1 - v,1 - v, - v,0,u)
          + G(1 - v,1 - v,u)\,\zeta_2
          + G(1 - v,1 - v,u)\,H(0,v)\,i\,\pi \\ &
          - G(1 - v,1 - v,u)\,H(0,0,v)
          + G(1 - v,1 - v,u)\,H(1,0,v)
          + G(1 - v,1 - v,0,u)\,H(0,v) \\ &
          + G(1 - v,u)\,\zeta_3
          + G(1 - v,u)\,H(0,v)\,\zeta_2
          + G(1 - v,u)\,H(0,0,v)\,i\,\pi \\ &
          - G(1 - v,u)\,H(0,0,0,v)
          + G(1 - v,u)\,H(0,1,0,v)
          + G(1 - v,u)\,H(1,0,0,v) \\ &
          - G(1 - v,0, - v,u)\,i\,\pi
          + G(1 - v,0, - v,u)\,H(0,v)
          - G(1 - v,0, - v,0,u) \\ &
          + G(1 - v,0,u)\,H(0,0,v)
          + G(1,1 - v, - v,u)\,i\,\pi
          - G(1,1 - v, - v,u)\,H(0,v) \\ &
          + G(1,1 - v, - v,0,u)
          + G(1,1 - v,u)\,\zeta_2
          + G(1,1 - v,u)\,H(0,v)\,i\,\pi \\ &
          - G(1,1 - v,u)\,H(0,0,v)
          + G(1,1 - v,u)\,H(1,0,v)
          + G(1,1 - v,0,u)\,H(0,v) \\ &
          + G(1,u)\,\zeta_3
          - 2\,G(1,u)\,i\,\pi\,\zeta_2
          - G(1,u)\,H(0,v)\,\zeta_2
          - G(1,u)\,H(0,0,v)\,i\,\pi \\ &
          + G(1,u)\,H(0,0,0,v)
          - G(1,u)\,H(0,1,0,v)
          - G(1,u)\,H(1,v)\,\zeta_2  
          - G(1,u)\,H(1,0,v)\,i\,\pi \\ &
          + G(1,u)\,H(1,0,0,v)
          - G(1,u)\,H(1,1,0,v) 
          + G(1,0, - v,u)\,i\,\pi
          - G(1,0, - v,u)\,H(0,v)\\ &
          + G(1,0, - v,0,u) 
          - 2\,G(1,0,u)\,\zeta_2
          - G(1,0,u)\,H(0,0,v)
          - G(1,0,u)\,H(1,0,v)
          \Big]\\
       &+ \frac{1}{16\,(1-u)\,(u+v)^2}\,\Big( - v - v^2 + 2\,v^3 - 4\,u\,v + 5\,u\,
      v^2 - 4\,u^2 + 7\,u^2\,v + 4\,u^3\, \Big) \, \Big[
            i\,\pi
          + G(0,u)
          \Big]\\
       &+ \frac{1}{144\,(1-u)^2\,(1-u-v)}\,\Big(313 - 331\,v + 18\,v^2 - 975\,u + 
      716\,u\,v - 54\,u\,v^2 + 1011\,u^2 \\ &- 385\,u^2\,v - 349\,u^3\, \Big) \, \Big[
            G(0,u)\,H(0,v)
          + H(0,v)\,i\,\pi
          \Big]\\
       &+ \frac{1}{16\,u\,(u+v)}\,\Big( - 8\,v + 8\,v^2 - 9\,u + 25\,u\,v + 17\,u^2 \, \Big)
          \, \\
       &+ \frac{1}{144\,u\,(1-u)^2\,(1-u-v)}\,\Big(36 - 108\,v + 108\,v^2 - 36\,v^3
       + 169\,u + 29\,u\,v - 270\,u\,v^2 \\ &+ 72\,u\,v^3 - 759\,u^2 + 320\,u^2\,v + 126\,u^2
      \,v^2 + 867\,u^3 - 241\,u^3\,v - 313\,u^4\, \Big) \, 
           H(1,0,v)
          \\
       &+ \frac{1}{144\,u\,(1-u)^2\,(1-u-v)}\,\Big(72 - 216\,v + 216\,v^2 - 72\,v^3
       + 25\,u + 389\,u\,v - 558\,u\,v^2 \\ &+ 144\,u\,v^3 - 543\,u^2 - 76\,u^2\,v + 306\,u^2
      \,v^2 + 723\,u^3 - 97\,u^3\,v - 277\,u^4\, \Big) \,
           \zeta_2
          \\
       &+ \frac{1}{144\,u\,(1-u)^2\,(1-u-v)}\,\Big( - 36\,v + 72\,v^2 - 36\,v^3 - 
      313\,u + 475\,u\,v - 234\,u\,v^2 \\ &+ 72\,u\,v^3 + 975\,u^2 - 896\,u^2\,v + 198\,u^2\,
      v^2 - 1011\,u^3 + 457\,u^3\,v + 349\,u^4\, \Big) \, \Big[
          - G( - v,u)\,i\,\pi\\ &
          + G( - v,u)\,H(0,v)
          - G( - v,0,u)
          + H(0,0,v)
          \Big]\\
       &+ \frac{1}{8\,u\,(1-u)^3}\,\Big(4 - 6\,v + 2\,v^2 - 10\,u + 16\,u\,v - 5\,u\,
      v^2 + 6\,u^2 - 16\,u^2\,v + 5\,u^2\,v^2 \\ &+ 2\,u^3 + 6\,u^3\,v - 2\,u^4\, \Big) \, \Big[
            G(1 - v, - v,u)\,i\,\pi
          - G(1 - v, - v,u)\,H(0,v)
          + G(1 - v, - v,0,u)\\ &
          + G(1 - v,u)\,\zeta_2
          + G(1 - v,u)\,H(0,v)\,i\,\pi
          - G(1 - v,u)\,H(0,0,v)
          + G(1 - v,u)\,H(1,0,v)\\ &
          + G(1 - v,0,u)\,H(0,v)
          \Big]\\
       &- \frac{1}{12\,u\,(1-u)^3}\,\Big(6 - 12\,v + 6\,v^2 - 4\,u + 30\,u\,v - 15\,u
      \,v^2 - 24\,u^2 - 30\,u^2\,v + 15\,u^2\,v^2 \\ &+ 36\,u^3 + 12\,u^3\,v - 14\,u^4\, \Big)
       \, 
           G(0,u)\,\zeta_2
          \\
       &+ \frac{1}{24\,u\,(1-u)^3}\,\Big(6 - 12\,v + 6\,v^2 + 7\,u + 30\,u\,v - 15\,u
      \,v^2 - 57\,u^2 - 30\,u^2\,v + 15\,u^2\,v^2 \\ &+ 69\,u^3 + 12\,u^3\,v - 25\,u^4\, \Big)
       \, \Big[
          - G(0,u)\,H(1,0,v)
          - H(1,v)\,\zeta_2
          - H(1,1,0,v)
          \Big]\\
       &+ \frac{1}{24\,u\,(1-u)^3}\,\Big(12 - 18\,v + 6\,v^2 - 8\,u + 48\,u\,v - 15\,
      u\,v^2 - 48\,u^2 - 48\,u^2\,v + 15\,u^2\,v^2 \\ &+ 72\,u^3 + 18\,u^3\,v - 28\,u^4 \, \Big)
        \,\zeta_3 \\
       &+ \frac{1}{24\,u\,(1-u)^3}\,\Big(12 - 18\,v + 6\,v^2 + 14\,u + 48\,u\,v - 15
      \,u\,v^2 - 114\,u^2 - 48\,u^2\,v + 15\,u^2\,v^2 \\ &+ 138\,u^3 + 18\,u^3\,v - 50\,u^4, \Big)
           \, H(1,0,0,v) \\
       &+ \frac{1}{24\,u\,(1-u)^3}\,\Big( - 6\,v + 6\,v^2 - 44\,u + 12\,u\,v - 15\,u\,
      v^2 + 132\,u^2 - 12\,u^2\,v + 15\,u^2\,v^2 \\ &- 132\,u^3 + 6\,u^3\,v + 44\,u^4\, \Big)
       \, \Big[
          - G(0,u)\,H(0,0,v)
          + H(0,0,0,v)
          \Big]\\
       &+ \frac{1}{24\,u\,(1-u)^3}\,\Big( - 6\,v + 6\,v^2 - 22\,u + 12\,u\,v - 15\,u\,
      v^2 + 66\,u^2 - 12\,u^2\,v + 15\,u^2\,v^2 \\ &- 66\,u^3 + 6\,u^3\,v + 22\,u^4\, \Big)
       \, \Big[
          - H(0,0,v)\,i\,\pi
          - H(0,1,0,v)
          \Big]\\
       &+ \frac{1}{24\,u\,(1-u)^3}\,\Big(6\,v - 6\,v^2 - 22\,u - 12\,u\,v + 15\,u\,v^2
       + 66\,u^2 + 12\,u^2\,v - 15\,u^2\,v^2 \\ &- 66\,u^3 - 6\,u^3\,v + 22\,u^4\, \Big) \,
          H(0,v)\,\zeta_2 \, ,\\ \\
   A_\alpha^{(2)} &=
        - \frac{11}{16} \, 
          H(0,v)\,\zeta_2
        + \frac{49}{32} \,
            \zeta_4
        + \frac{7}{4} \, \Big[
            i\,\pi\,\zeta_3
          + H(0,v)\,\zeta_3
          \Big]\\
       &+ \frac{(1-v)^2}{4\,u^2} \, \Bigg[
            \frac{13}{6}\,G(1,u)\,H(0,v)\,i\,\pi
          - \frac{35}{6}\,G(1,u)\,H(0,1,v)
          + \frac{22}{3}\,G(1 - v,u)\,H(0,0,v)\\ &
          - \frac{22}{3}\,G(1,u)\,H(0,0,v)
          + 2\,G(1 - v,1 - v,u)\,\zeta_2
          + 2\,G(1 - v,1 - v,u)\,H(0,v)\,i\,\pi\\ &
          - 2\,G(1 - v,1 - v,u)\,H(0,1,v)
          - 2\,G(1 - v,1 - v,1,1 - v,u)
          - 2\,G(1 - v,1 - v,1,u)\,i\,\pi\\ &
          + 2\,G(1 - v,1 - v,1,u)\,H(0,v)
          + 2\,G(1 - v,1 - v,1,u)\,H(1,v)
          + 2\,G(1 - v,0,1 - v,u)\,H(0,v)\\ &
          + 2\,G(1 - v,0,1,1 - v,u)
          + 2\,G(1 - v,0,1,u)\,i\,\pi
          - 2\,G(1 - v,0,1,u)\,H(0,v)\\ &
          - 2\,G(1 - v,0,1,u)\,H(1,v)
          + G(1 - v,1,1 - v,u)\,H(0,v)
          - 4\,G(1 - v,1,u)\,\zeta_2\\ &
          - G(1 - v,1,u)\,H(0,v)\,i\,\pi
          + G(1 - v,1,u)\,H(0,1,v)
          - G(1 - v,1,u)\,H(1,0,v)\\ &
          + 2\,G(1 - v,1,1,1 - v,u)
          + 2\,G(1 - v,1,1,u)\,i\,\pi
          - 2\,G(1 - v,1,1,u)\,H(0,v)\\ &
          - 2\,G(1 - v,1,1,u)\,H(1,v)
          + 2\,G(1,1 - v,1 - v,u)\,H(0,v)
          - 4\,G(1,1 - v,u)\,\zeta_2\\ &
          - G(1,1 - v,u)\,H(0,v)\,i\,\pi
          + G(1,1 - v,u)\,H(0,1,v)
          - G(1,1 - v,u)\,H(1,0,v)\\ &
          + 3\,G(1,1 - v,1,1 - v,u)
          + 3\,G(1,1 - v,1,u)\,i\,\pi
          - 3\,G(1,1 - v,1,u)\,H(0,v)\\ &
          - 3\,G(1,1 - v,1,u)\,H(1,v)
          - 3\,G(1,u)\,\zeta_3
          - G(1,u)\,i\,\pi\,\zeta_2
          + 4\,G(1,u)\,H(1,v)\,\zeta_2\\ &
          + G(1,u)\,H(1,0,v)\,i\,\pi
          - G(1,u)\,H(1,0,1,v)
          + G(1,u)\,H(1,1,0,v)\\ &
          - G(1,0,1 - v,u)\,H(0,v)
          - G(1,0,1,1 - v,u)
          - G(1,0,1,u)\,i\,\pi
          + G(1,0,1,u)\,H(0,v)\\ &
          + G(1,0,1,u)\,H(1,v)
          + 2\,G(1,1,1 - v,1 - v,u)
          + 2\,G(1,1,1 - v,u)\,i\,\pi\\ &
          - 2\,G(1,1,1 - v,u)\,H(0,v)
          - 2\,G(1,1,1 - v,u)\,H(1,v)
          - 2\,G(1,1,u)\,\zeta_2\\ &
          - 2\,G(1,1,u)\,H(1,v)\,i\,\pi
          + 2\,G(1,1,u)\,H(1,0,v)
          + 2\,G(1,1,u)\,H(1,1,v)\\ &
          - 2\,G(1,1,1,1 - v,u)
          - 2\,G(1,1,1,u)\,i\,\pi
          + 2\,G(1,1,1,u)\,H(0,v)
          + 2\,G(1,1,1,u)\,H(1,v)
          \Bigg]\\
       &+ \frac{11\,v}{12\,u\,(1-u)^2}\,\Big(2 - 2\,v - 2\,u + 3\,u\,v\, \Big) \, 
          H(0,0,v)
          \\
       &+ \frac{(1-u-v)}{8\,u\,(1-u)^2}\,\Big(2 - 2\,v - u + 3\,u\,v - u^2\, \Big)
       \, \Big[
          - G(1 - v,u)\,H(1,0,v)
          + H(1,0,v)\,i\,\pi \\ &
          - H(1,0,1,v)
          + H(1,1,0,v)
          \Big]\\
       &+ \frac{(1-u-v)}{16\,u\,(1-u)^2}\,\Big(6 - 6\,v - 3\,u + 5\,u\,v - 3\,u^2 \, \Big)
            \, \Big[
            G(1 - v,u)\,i\,\pi
          - H(1,v)\,i\,\pi
          \Big]\\
       &- \frac{(1-u-v)}{24\,u\,(1-u)^2}\,\Big(22 - 22\,v - 11\,u + 39\,u\,v - 11\,
      u^2\, \Big) \, 
           H(1,0,v)
          \\
       &+ \frac{(1-u-v)}{48\,u\,(1-u)^2}\,\Big(26 - 26\,v - 13\,u + 51\,u\,v - 13\,
      u^2\, \Big) \, \Big[
          - G(1 - v,1 - v,u) \\ &
          + G(1 - v,u)\,H(1,v)
          - H(1,1,v)
          \Big]\\
       &+ \frac{(1-u-v)}{288\,u\,(1-u)^2}\,\Big( - 554 + 288\,\zeta_2 + 554\,v - 
      288\,v\,\zeta_2 + 433\,u - 144\,u\,\zeta_2 - 909\,u\,v \\ &+ 432\,u\,v\,\zeta_2 + 121\,u^2 - 
      144\,u^2\,\zeta_2\, \Big) \, 
           H(1,v)
          \\
       &+ \frac{(1-u-v)}{288\,u\,(1-u)^2}\,\Big(554 - 554\,v - 433\,u + 909\,u\,v
       - 121\,u^2\, \Big) \,
           G(1 - v,u)
          \\
       &+ \frac{(1-u-v)}{8\,u^2\,(1-u)^2}\,\Big(3 - 3\,v - u + 2\,u\,v - u^2 + 3\,
      u^2\,v - u^3\, \Big) \, \Big[
            G(1,1 - v,u)\,i\,\pi
          - G(1,u)\,H(1,v)\,i\,\pi
          \Big]\\
       &+ \frac{1}{10368\,u\,(1-u)}\,\Big(5184 - 5184\,v - 87239\,u - 4788\,u\,v + 
      82055\,u^2\, \Big) \\
       &+ \frac{1}{8\,u\,(1-u)^2}\,\Big(2 - 8\,v + 6\,v^2 - 3\,u + 10\,u\,v - 9\,u\,
      v^2 - 2\,u^2\,v + u^3\, \Big) \, \Big[
            G(1 - v,1,1 - v,u)\\ &
          + G(1 - v,1,u)\,i\,\pi 
          - G(1 - v,1,u)\,H(0,v)
          - G(1 - v,1,u)\,H(1,v)
          \Big]\\
       &+ \frac{1}{8\,u\,(1-u)^2}\,\Big(2 - 2\,v^2 - 3\,u + 2\,u\,v + 3\,u\,v^2 - 2\,
      u^2\,v + u^3\, \Big) \, \Big[
            G(0,1 - v,u)\,H(0,v)\\ &
          + G(0,1,1 - v,u)
          + G(0,1,u)\,i\,\pi
          - G(0,1,u)\,H(0,v)
          - G(0,1,u)\,H(1,v)
          \Big]\\
       &- \frac{1}{48\,u\,(1-u)^2}\,\Big(44 - 114\,v + 70\,v^2 - 66\,u + 158\,u\,v
       - 105\,u\,v^2 - 44\,u^2\,v + 22\,u^3\, \Big) \,
           H(0,1,v) \\
       &+ \frac{1}{48\,u\,(1-u)^2}\,\Big(44 - 70\,v + 26\,v^2 - 66\,u + 114\,u\,v - 
      39\,u\,v^2 - 44\,u^2\,v + 22\,u^3\, \Big) \, 
           H(0,v)\,i\,\pi \\
       &- \frac{1}{144\,u\,(1-u)^2}\,\Big(480\,\zeta_2 + 108\,\zeta_3 - 882\,v\,\zeta_2
       - 216\,v\,\zeta_3 + 402\,v^2\,\zeta_2 + 108\,v^2\,\zeta_3 - 660\,u\,\zeta_2 \\ &- 551\,u\,
      \zeta_3 + 1326\,u\,v\,\zeta_2 + 324\,u\,v\,\zeta_3 - 567\,u\,v^2\,\zeta_2 - 162\,u\,v^2\,
      \zeta_3 - 120\,u^2\,\zeta_2 + 778\,u^2\,\zeta_3 \\ &- 444\,u^2\,v\,\zeta_2 - 108\,u^2\,v\,
      \zeta_3 + 300\,u^3\,\zeta_2 - 335\,u^3\,\zeta_3\, \Big) \, \\
       &- \frac{1}{864\,u\,(1-u)^2}\,\Big( - 1662 + 216\,\zeta_2 + 3324\,v - 432\,v\,
      \zeta_2 - 1662\,v^2 + 216\,v^2\,\zeta_2 + 5324\,u \\ &+ 270\,u\,\zeta_2 - 5292\,u\,v + 648
      \,u\,v\,\zeta_2 + 2727\,u\,v^2 - 324\,u\,v^2\,\zeta_2 - 5662\,u^2 - 1188\,u^2\,\zeta_2 + 
      1968\,u^2\,v \\ &- 216\,u^2\,v\,\zeta_2 + 2000\,u^3 + 702\,u^3\,\zeta_2\, \Big) \, 
           i\,\pi
          \\
       &- \frac{1}{864\,u\,(1-u)^2}\,\Big( - 1662\,v + 1662\,v^2 + 2363\,u + 2526\,
      u\,v - 2727\,u\,v^2 - 4726\,u^2 - 864\,u^2\,v \\ &+ 2363\,u^3\, \Big) \, 
          H(0,v)
          \\
       &+ \frac{1}{8\,u^2\,(1-u)^2}\,\Big(3 - 6\,v + 3\,v^2 - 6\,u + 14\,u\,v - 8\,u\,
      v^2 + 3\,u^2 - 8\,u^2\,v + 6\,u^2\,v^2\, \Big) \\ & \times\, 
            G(1 - v,u)\,H(0,v)\,i\,\pi
          \\
       &+ \frac{1}{24\,u^2\,(1-u)^2}\,\Big(13 - 26\,v + 13\,v^2 - 32\,u + 70\,u\,v
       - 38\,u\,v^2 + 22\,u^2 - 50\,u^2\,v + 31\,u^2\,v^2 \\ &+ 6\,u^3\,v - 3\,u^4\, \Big)
       \, \Big[
          - G(1 - v,1 - v,u)\,H(0,v)
          - G(1,1 - v,1 - v,u)
          + G(1,1 - v,u)\,H(1,v)\\ &
          - G(1,u)\,H(1,1,v)
          \Big]\\
       &+ \frac{1}{24\,u^2\,(1-u)^2}\,\Big(13 - 26\,v + 13\,v^2 - 26\,u + 46\,u\,v
       - 20\,u\,v^2 + 13\,u^2 - 20\,u^2\,v + 4\,u^2\,v^2\, \Big) \,\\ &\times 
            G(1 - v,u)\,H(0,1,v)
          \\
       &+ \frac{1}{24\,u^2\,(1-u)^2}\,\Big(13 - 26\,v + 13\,v^2 - 20\,u + 34\,u\,v
       - 14\,u\,v^2 + 4\,u^2 - 2\,u^2\,v - 5\,u^2\,v^2 - 6\,u^3\,v \\ &+ 3\,u^4\, \Big) \, \Big[
          - G(1,1,1 - v,u)
          - G(1,1,u)\,i\,\pi
          + G(1,1,u)\,H(0,v)
          + G(1,1,u)\,H(1,v)
          \Big]\\
       &- \frac{1}{24\,u^2\,(1-u)^2}\,\Big(13 - 26\,v + 13\,v^2 - 8\,u + 10\,u\,v - 
      2\,u\,v^2 - 14\,u^2 + 34\,u^2\,v - 23\,u^2\,v^2 \\ &- 18\,u^3\,v + 9\,u^4\, \Big) \, 
          G(1 - v,u)\,\zeta_2\\
       &+ \frac{1}{24\,u^2\,(1-u)^2}\,\Big(22 - 44\,v + 22\,v^2 - 50\,u + 106\,u\,v
       - 56\,u\,v^2 + 31\,u^2 - 68\,u^2\,v + 40\,u^2\,v^2 \\ &+ 6\,u^3\,v - 3\,u^4\, \Big)
       \, \Big[
            G(1,1 - v,u)\,H(0,v)
          - G(1,u)\,H(1,0,v)
          \Big]\\
       &- \frac{1}{24\,u^2\,(1-u)^2}\,\Big(67 - 134\,v + 67\,v^2 - 128\,u + 250\,u\,
      v - 122\,u\,v^2 + 58\,u^2 - 110\,u^2\,v + 49\,u^2\,v^2\\ & - 6\,u^3\,v + 3\,u^4\, \Big)
       \,
           G(1,u)\,\zeta_2
         \\
       &+ \frac{1}{144\,u^2\,(1-u)^2}\,\Big(277 - 554\,v + 277\,v^2 - 650\,u + 
      1264\,u\,v - 614\,u\,v^2 + 430\,u^2 - 770\,u^2\,v \\ &+ 340\,u^2\,v^2 - 18\,u^3 + 60\,
      u^3\,v - 39\,u^4\, \Big) \, \Big[
            G(1,1 - v,u)
          + G(1,u)\,i\,\pi
          - G(1,u)\,H(0,v)\\ &
          - G(1,u)\,H(1,v)
          \Big]\\
       &+ \frac{1}{144\,u^2\,(1-u)^2}\,\Big(277 - 554\,v + 277\,v^2 - 518\,u + 922
      \,u\,v - 404\,u\,v^2 + 232\,u^2 - 296\,u^2\,v \\ &+ 25\,u^2\,v^2 - 18\,u^3 - 72\,u^3\,v
       + 27\,u^4\, \Big) \, 
           G(1 - v,u)\,H(0,v) \, ,
\end{align*}
\begin{align}
   A_\alpha^{(3)} &=
         \frac{(1-u-v)}{(1-v)} \, \Bigg[
          - \frac{5}{12}\,\zeta_2
          - \frac{11}{16}\,i\,\pi\,\zeta_2
          - \frac{11}{16}\,H(0,v)\,\zeta_2
          + \frac{49}{32}\,\zeta_4
          + \frac{7}{4}\,i\,\pi\,\zeta_3
          + \frac{7}{4}\,H(0,v)\,\zeta_3
          + \frac{389}{144}\,\zeta_3\nonumber \\ &
          - \frac{2759}{864}\,i\,\pi
          - \frac{2759}{864}\,H(0,v)
          - \frac{81659}{10368}
          \Bigg],\nonumber \\ \nonumber \\
   A_\alpha^{(4)} &= 0.
\end{align}

\begin{align*}
   A_\beta^{(1)} &=
         \frac{1}{4} \, \Big[
          - G( - v,1 - v, - v,u)\,i\,\pi
          + G( - v,1 - v, - v,u)\,H(0,v)
          - G( - v,1 - v, - v,0,u)\\ &
          - G( - v,1 - v,u)\,\zeta_2
          - G( - v,1 - v,u)\,H(0,v)\,i\,\pi
          + G( - v,1 - v,u)\,H(0,0,v)\\ &
          - G( - v,1 - v,u)\,H(1,0,v)
          - G( - v,1 - v,0,u)\,H(0,v)
          - 2\,G( - v, - v, - v,u)\,i\,\pi\\ &
          + 2\,G( - v, - v, - v,u)\,H(0,v)
          - 2\,G( - v, - v, - v,0,u)
          - 4\,G( - v, - v,u)\,\zeta_2\\ &
          - 2\,G( - v, - v,u)\,H(0,v)\,i\,\pi
          + 2\,G( - v, - v,u)\,H(0,0,v)
          + 2\,G( - v, - v,0,u)\,i\,\pi\\ &
          - 2\,G( - v, - v,0,u)\,H(0,v)
          + 2\,G( - v, - v,0,0,u)
          - 2\,G( - v,u)\,\zeta_3
          - G( - v,u)\,i\,\pi\,\zeta_2\\ &
          - 3\,G( - v,u)\,H(0,0,v)\,i\,\pi
          + 3\,G( - v,u)\,H(0,0,0,v)
          - G( - v,u)\,H(0,1,0,v)
          + G( - v,u)\,H(1,v)\,\zeta_2\\ &
          + G( - v,u)\,H(1,0,v)\,i\,\pi
          - G( - v,u)\,H(1,0,0,v)
          + G( - v,u)\,H(1,1,0,v)\\ &
          + 3\,G( - v,0, - v,u)\,i\,\pi
          - 3\,G( - v,0, - v,u)\,H(0,v)
          + 3\,G( - v,0, - v,0,u)
          - G( - v,0,u)\,\zeta_2\\ &
          + 2\,G( - v,0,u)\,H(0,v)\,i\,\pi
          - 3\,G( - v,0,u)\,H(0,0,v)
          + G( - v,0,u)\,H(1,0,v)\\ &
          + 2\,G( - v,0,0,u)\,H(0,v)
          + 2\,G(0,1 - v, - v,u)\,i\,\pi
          - 2\,G(0,1 - v, - v,u)\,H(0,v)\\ &
          + 2\,G(0,1 - v, - v,0,u)
          + 2\,G(0,1 - v,u)\,\zeta_2
          + 2\,G(0,1 - v,u)\,H(0,v)\,i\,\pi\\ &
          - 2\,G(0,1 - v,u)\,H(0,0,v)
          + 2\,G(0,1 - v,u)\,H(1,0,v)
          + 2\,G(0,1 - v,0,u)\,H(0,v)\\ &
          + 2\,G(0, - v, - v,u)\,i\,\pi
          - 2\,G(0, - v, - v,u)\,H(0,v)
          + 2\,G(0, - v, - v,0,u)
          - 2\,G(0, - v,u)\,\zeta_2\\ &
          + G(0, - v,u)\,H(0,v)\,i\,\pi
          - 2\,G(0, - v,u)\,H(0,0,v)
          + G(0, - v,0,u)\,H(0,v)
          - G(0,u)\,H(0,v)\,\zeta_2\\ &
          + 2\,G(0,u)\,H(0,0,v)\,i\,\pi
          - 3\,G(0,u)\,H(0,0,0,v)
          + G(0,u)\,H(0,1,0,v)\\ &
          + 2\,G(0,u)\,H(1,0,0,v)
          - 2\,G(0,0, - v,u)\,i\,\pi
          + 2\,G(0,0, - v,u)\,H(0,v)\\ &
          - 2\,G(0,0, - v,0,u)
          + 2\,G(0,0,u)\,H(0,0,v)
          - H(0,v)\,i\,\pi\,\zeta_2\\ &
          - 3\,H(0,0,0,v)\,i\,\pi
          + 3\,H(0,0,0,0,v)
          - H(0,0,1,0,v)
          + H(0,1,v)\,\zeta_2\\ &
          + H(0,1,0,v)\,i\,\pi
          - H(0,1,0,0,v)
          + H(0,1,1,0,v)
          + 2\,H(1,v)\,\zeta_3
          + 2\,H(1,0,v)\,\zeta_2\\ &
          + 2\,H(1,0,0,v)\,i\,\pi
          - 2\,H(1,0,0,0,v)
          + 2\,H(1,0,1,0,v)
          + 2\,H(1,1,0,0,v)
          \Big]\\
       &+ \frac{11}{12} \, \Big[
          - G( - v, - v,u)\,i\,\pi
          + G( - v, - v,u)\,H(0,v)
          - G( - v, - v,0,u)
          - G( - v,u)\,H(0,0,v)\\ &
          + G( - v,0,u)\,H(0,v)
          - G( - v,0,0,u)
          - G(0,0,u)\,H(0,v)
          \Big]\\
       &+ \frac{39}{32} \, \zeta_4
       + \frac{5}{4} \, \Big[
          - G( - v,u)\,H(0,v)\,\zeta_2
          + H(0,v)\,\zeta_3
          - H(0,0,v)\,\zeta_2
          \Big]
        + \frac{7}{4} \, 
           i\,\pi\,\zeta_3
       - \frac{11}{6} \,
          G( - v,u)\,\zeta_2 \\
       &+ \frac{v}{8\,u\,(1-u)^2\,(1-v)^2}\,\Big(2 - 4\,v + 2\,v^2 - 8\,u + 13\,u\,v
       - 5\,u\,v^2 + 12\,u^2 - 16\,u^2\,v + 5\,u^2\,v^2 \\ &- 5\,u^3 + 3\,u^3\,v + u^4\, \Big)
       \, \Big[
          - G(0, - v,u)\,i\,\pi
          + G(0, - v,u)\,H(0,v)
          - G(0, - v,0,u)
          \Big]\\
       &+ \frac{(1-u)}{4\,u^2}\,\Big(1 - v + u\, \Big) \, \Big[
            G(1 - v,1 - v, - v,u)\,i\,\pi
          - G(1 - v,1 - v, - v,u)\,H(0,v)\\ &
          + G(1 - v,1 - v, - v,0,u)
          + G(1 - v,1 - v,u)\,\zeta_2
          + G(1 - v,1 - v,u)\,H(0,v)\,i\,\pi\\ &
          - G(1 - v,1 - v,u)\,H(0,0,v)
          + G(1 - v,1 - v,u)\,H(1,0,v)
          + G(1 - v,1 - v,0,u)\,H(0,v)\\ &
          + G(1 - v,u)\,\zeta_3
          + G(1 - v,u)\,H(0,v)\,\zeta_2
          + G(1 - v,u)\,H(0,0,v)\,i\,\pi
          - G(1 - v,u)\,H(0,0,0,v)\\ &
          + G(1 - v,u)\,H(0,1,0,v)
          + G(1 - v,u)\,H(1,0,0,v)
          - G(1 - v,0, - v,u)\,i\,\pi\\ &
          + G(1 - v,0, - v,u)\,H(0,v)
          - G(1 - v,0, - v,0,u)
          + G(1 - v,0,u)\,H(0,0,v)\\ &
          + G(1,1 - v, - v,u)\,i\,\pi
          - G(1,1 - v, - v,u)\,H(0,v)
          + G(1,1 - v, - v,0,u)
          + G(1,1 - v,u)\,\zeta_2\\ &
          + G(1,1 - v,u)\,H(0,v)\,i\,\pi
          - G(1,1 - v,u)\,H(0,0,v)
          + G(1,1 - v,u)\,H(1,0,v)\\ &
          + G(1,1 - v,0,u)\,H(0,v)
          + G(1,u)\,\zeta_3
          - 2\,G(1,u)\,i\,\pi\,\zeta_2
          - G(1,u)\,H(0,v)\,\zeta_2\\ &
          - G(1,u)\,H(0,0,v)\,i\,\pi
          + G(1,u)\,H(0,0,0,v)
          - G(1,u)\,H(0,1,0,v)
          - G(1,u)\,H(1,v)\,\zeta_2\\ &
          - G(1,u)\,H(1,0,v)\,i\,\pi
          + G(1,u)\,H(1,0,0,v)
          - G(1,u)\,H(1,1,0,v)
          + G(1,0, - v,u)\,i\,\pi\\ &
          - G(1,0, - v,u)\,H(0,v)
          + G(1,0, - v,0,u)
          - 2\,G(1,0,u)\,\zeta_2
          - G(1,0,u)\,H(0,0,v)\\ &
          - G(1,0,u)\,H(1,0,v)
          \Big]\\
       &- \frac{1}{48\,(1-v)\,(u+v)^2}\,\Big(3\,v - 19\,v^2 + 16\,v^3 - 17\,u\,v + 
      36\,u\,v^2 - u^2 + 39\,u^2\,v + 19\,u^3\, \Big) \, G(0,u) \\
       &+ \frac{1}{864\,(1-v)\,(u+v)^2}\,\Big( - 54\,v - 2813\,v^2 + 2867\,v^3 - 
      6004\,u\,v + 6058\,u\,v^2 - 3137\,u^2 \\ &+ 3245\,u^2\,v + 54\,u^3\, \Big) \,
          i\,\pi \\
       &+ \frac{1}{864\,(1-v)^2\,(u+v)^2}\,\Big(54\,v - 3533\,v^2 + 4537\,v^3 - 
      1058\,v^4 - 6580\,u\,v + 8822\,u\,v^2 \\ &- 1753\,u\,v^3 - 3155\,u^2 + 4735\,u^2\,v - 
      602\,u^2\,v^2 + 396\,u^3 + 93\,u^3\,v\, \Big) \, 
           H(0,v)
          \\
       &+ \frac{1}{144\,(1-u)\,(1-v)^2\,(1-u-v)}\,\Big(295 - 912\,v + 939\,v^2 - 
      322\,v^3 - 626\,u + 1700\,u\,v \\ &- 1432\,u\,v^2 + 358\,u\,v^3 + 313\,u^2 - 761\,u^2\,
      v + 421\,u^2\,v^2 + 18\,u^3 + 9\,u^3\,v\, \Big) \,
            H(0,v)\,i\,\pi \\
       &+ \frac{1}{144\,(1-u)\,(1-v)^2\,(1-u-v)}\,\Big(295 - 714\,v + 543\,v^2 - 
      124\,v^3 - 626\,u + 1238\,u\,v \\ &- 772\,u\,v^2 + 160\,u\,v^3 + 313\,u^2 - 431\,u^2\,v
       + 157\,u^2\,v^2 + 18\,u^3 - 57\,u^3\,v\, \Big) \,
           G(0,u)\,H(0,v) \\
       &+ \frac{1}{10368\,u\,(1-v)\,(u+v)}\,\Big( - 5184\,v + 5184\,v^2 - 5832\,u
       - 79967\,u\,v + 85799\,u\,v^2 \\ &- 85799\,u^2 + 95771\,u^2\,v + 9972\,u^3\, \Big)
       \,\\
       &+ \frac{1}{144\,u\,(1-u)\,(1-v)\,(1-u-v)}\,\Big(36\,v - 72\,v^2 + 36\,v^3 + 
      295\,u - 644\,u\,v + 421\,u\,v^2 \\ &- 72\,u\,v^3 - 626\,u^2 + 957\,u^2\,v - 349\,u^2\,
      v^2 + 313\,u^3 - 313\,u^3\,v + 18\,u^4\, \Big) \, \Big[
           G( - v,u)\,i\,\pi\\ &
          - G( - v,u)\,H(0,v)
          + G( - v,0,u)
          \Big]\\
       &+ \frac{1}{144\,u\,(1-u)\,(1-v)^2\,(1-u-v)}\,\Big(36 - 144\,v + 216\,v^2 - 
      144\,v^3 + 36\,v^4 + 187\,u - 318\,u\,v \\ &+ 3\,u\,v^2 + 200\,u\,v^3 - 72\,u\,v^4 - 
      518\,u^2 + 914\,u^2\,v - 448\,u^2\,v^2 + 52\,u^2\,v^3 + 277\,u^3 - 359\,u^3\,v \\ &+ 
      121\,u^3\,v^2 + 18\,u^4 - 57\,u^4\,v\, \Big) \, 
           H(1,0,v)
          \\
       &+ \frac{1}{144\,u\,(1-u)\,(1-v)^2\,(1-u-v)}\,\Big(72 - 288\,v + 432\,v^2 - 
      288\,v^3 + 72\,v^4 + 19\,u + 258\,u\,v \\ &- 717\,u\,v^2 + 584\,u\,v^3 - 144\,u\,v^4 - 
      290\,u^2 + 290\,u^2\,v + 116\,u^2\,v^2 - 116\,u^2\,v^3 + 181\,u^3 \\ &- 167\,u^3\,v + 
      25\,u^3\,v^2 + 18\,u^4 - 57\,u^4\,v\, \Big) \, 
           \zeta_2
          \\
       &+ \frac{1}{144\,u\,(1-u)\,(1-v)^2\,(1-u-v)}\,\Big( - 36\,v + 108\,v^2 - 108
      \,v^3 + 36\,v^4 - 295\,u + 543\,u\,v \\ &- 273\,u\,v^2 + 97\,u\,v^3 - 72\,u\,v^4 + 626\,
      u^2 - 659\,u^2\,v - 14\,u^2\,v^2 + 47\,u^2\,v^3 - 313\,u^3 - 34\,u^3\,v \\ &+ 215\,u^3
      \,v^2 - 18\,u^4 + 150\,u^4\,v\, \Big) \, 
           H(0,0,v)
          \\
       &- \frac{1}{8\,u\,(1-u)^2}\,\Big(2 - 2\,v - 3\,u + 5\,u\,v - 5\,u^2\,v + u^3 \, \Big)
            \, 
           H(1,0,v)\,i\,\pi
          \\
       &- \frac{1}{12\,u\,(1-u)^2}\,\Big(6 - 6\,v + 2\,u + 15\,u\,v - 22\,u^2 - 15\,
      u^2\,v + 14\,u^3\, \Big) \, 
           G(0,u)\,\zeta_2
          \\
       &- \frac{1}{16\,u\,(1-u)^2}\,\Big(8 - 8\,v - u + 20\,u\,v - 22\,u^2 - 20\,u^2
      \,v + 15\,u^3\, \Big) \,
            i\,\pi\,\zeta_2
          \\
       &+ \frac{1}{24\,u\,(1-u)^2}\,\Big(6 - 6\,v + 13\,u + 15\,u\,v - 44\,u^2 - 15\,
      u^2\,v + 25\,u^3\, \Big) \, \Big[
          - G(0,u)\,H(1,0,v)\\ &
          - H(1,v)\,\zeta_2
          - H(1,1,0,v)
          \Big]\\
       &+ \frac{1}{8\,u\,(1-u)^2\,(1-v)^2}\,\Big(4 - 10\,v + 8\,v^2 - 2\,v^3 - 6\,u
       + 14\,u\,v - 13\,u\,v^2 + 5\,u\,v^3 + 2\,u^2\,v \\ &+ 4\,u^2\,v^2 - 5\,u^2\,v^3 + 2\,u^3
       - 9\,u^3\,v + 5\,u^3\,v^2 + u^4\,v\, \Big) \, \Big[
            G(1 - v, - v,u)\,i\,\pi\\ &
          - G(1 - v, - v,u)\,H(0,v)
          + G(1 - v, - v,0,u)
          + G(1 - v,u)\,\zeta_2
          + G(1 - v,u)\,H(0,v)\,i\,\pi\\ &
          - G(1 - v,u)\,H(0,0,v)
          + G(1 - v,u)\,H(1,0,v)
          + G(1 - v,0,u)\,H(0,v)
          \Big]\\
       &+ \frac{1}{24\,u\,(1-u)^2\,(1-v)^2}\,\Big(12 - 30\,v + 24\,v^2 - 6\,v^3 + 
      26\,u - 46\,u\,v + 5\,u\,v^2 + 15\,u\,v^3 - 88\,u^2 \\ &+ 182\,u^2\,v - 76\,u^2\,v^2 - 
      15\,u^2\,v^3 + 50\,u^3 - 115\,u^3\,v + 59\,u^3\,v^2 + 3\,u^4\,v\, \Big) \, H(1,0,0,v)\\
       &+ \frac{1}{24\,u\,(1-u)^2\,(1-v)^2}\,\Big(6\,v - 12\,v^2 + 6\,v^3 + 22\,u - 
      68\,u\,v + 61\,u\,v^2 - 15\,u\,v^3 - 44\,u^2 \\ &+ 124\,u^2\,v - 92\,u^2\,v^2 + 15\,u^2\,
      v^3 + 22\,u^3 - 59\,u^3\,v + 31\,u^3\,v^2 + 3\,u^4\,v\, \Big) \, \Big[
            H(0,0,v)\,i\,\pi\\ &
          + H(0,1,0,v)
          \Big]\\
       &+ \frac{1}{24\,u\,(1-u)^2\,(1-v)^2}\,\Big(6\,v - 12\,v^2 + 6\,v^3 + 44\,u - 
      112\,u\,v + 83\,u\,v^2 - 15\,u\,v^3 - 88\,u^2 \\ &+ 212\,u^2\,v - 136\,u^2\,v^2 + 15\,
      u^2\,v^3 + 44\,u^3 - 103\,u^3\,v + 53\,u^3\,v^2 + 3\,u^4\,v\, \Big) \, \Big[
            G(0,u)\,H(0,0,v)\\ &
          - H(0,0,0,v)
          \Big]\\
       &+ \frac{1}{48\,u\,(1-u)^2\,(1-v)^2}\,\Big(12\,v - 24\,v^2 + 12\,v^3 - 77\,u
       + 106\,u\,v + u\,v^2 - 30\,u\,v^3 + 154\,u^2 \\ &- 236\,u^2\,v + 58\,u^2\,v^2 + 30\,
      u^2\,v^3 - 77\,u^3 + 124\,u^3\,v - 59\,u^3\,v^2 + 6\,u^4\,v\, \Big) \, 
           H(0,v)\,\zeta_2
          \\
       &+ \frac{1}{144\,u\,(1-u)^2\,(1-v)^2}\,\Big(72 - 180\,v + 144\,v^2 - 36\,v^3
       + 413\,u - 790\,u\,v + 287\,u\,v^2 + 90\,u\,v^3 \\ &- 1042\,u^2 + 2120\,u^2\,v - 970\,
      u^2\,v^2 - 90\,u^2\,v^3 + 557\,u^3 - 1204\,u^3\,v + 611\,u^3\,v^2 + 18\,u^4\,v \, \Big)
            \,
          \zeta_3, \\ \\
   A_\beta^{(2)} &=
        \frac{11\,v}{12\,u\,(1-u)\,(1-v)^2}\,\Big(2 - 4\,v + 2\,v^2 + 3\,u\,v - 3\,u
      \,v^2 - 2\,u^2 + 3\,u^2\,v - u^3\, \Big) \, H(0,0,v)\\
       &+ \frac{1}{288\,u\,(1-v)}\,\Big(144 - 144\,v + 133\,u - 133\,u\,v + 277\,u^2
      \, \Big) \\
       &+ \frac{1}{8\,u\,(1-u)}\,\Big(2 - 2\,v - u + 3\,u\,v - u^2\, \Big) \, \Big[
          - 3\,\zeta_3
          - i\,\pi\,\zeta_2
          - G(1 - v,u)\,H(1,0,v)\\ &
          + 4\,H(1,v)\,\zeta_2
          + H(1,0,v)\,i\,\pi
          - H(1,0,1,v)
          + H(1,1,0,v)
          \Big]\\
       &+ \frac{1}{16\,u\,(1-u)}\,\Big(6 - 6\,v - 3\,u + 5\,u\,v - 3\,u^2\, \Big) \, \Big[
            G(1 - v,u)\,i\,\pi
          - H(1,v)\,i\,\pi
          \Big]\\
       &- \frac{1}{24\,u\,(1-u)}\,\Big(22 - 22\,v - 11\,u + 39\,u\,v - 11\,u^2\, \Big)
       \, 
           H(1,0,v)
          \\
       &+ \frac{1}{48\,u\,(1-u)}\,\Big(26 - 26\,v - 13\,u + 51\,u\,v - 13\,u^2\, \Big)
       \, \Big[
          - G(1 - v,1 - v,u)\\ &
          + G(1 - v,u)\,H(1,v)
          - H(1,1,v)
          \Big]\\
       &+ \frac{1}{288\,u\,(1-u)\,(1-v)}\,\Big(554 - 1108\,v + 554\,v^2 - 397\,u + 
      1306\,u\,v - 909\,u\,v^2 - 271\,u^2 \\ &+ 157\,u^2\,v + 114\,u^3\, \Big) \, \Big[
            G(1 - v,u)
          - H(1,v)
          \Big]\\
       &+ \frac{1}{288\,u\,(1-u)\,(1-v)}\,\Big(554 - 1108\,v + 554\,v^2 - 265\,u + 
      1174\,u\,v - 909\,u\,v^2 - 271\,u^2 \\ &+ 289\,u^2\,v - 18\,u^3\, \Big) \,
           i\,\pi\\
       &+ \frac{1}{8\,u\,(1-u)\,(1-v)^2}\,\Big(2 - 10\,v + 14\,v^2 - 6\,v^3 - u + 8
      \,u\,v - 16\,u\,v^2 + 9\,u\,v^3 - u^2 + 4\,u^2\,v \\ &- 4\,u^2\,v^2 + u^3\,v\, \Big) \, \Big[
            G(1 - v,1,1 - v,u)
          + G(1 - v,1,u)\,i\,\pi
          - G(1 - v,1,u)\,H(0,v)\\ &
          - G(1 - v,1,u)\,H(1,v)
          \Big]\\
       &+ \frac{1}{8\,u\,(1-u)\,(1-v)^2}\,\Big(2 - 2\,v - 2\,v^2 + 2\,v^3 - u + 2\,u
      \,v + 2\,u\,v^2 - 3\,u\,v^3 - u^2 + 2\,u^2\,v^2 \\ &- u^3\,v\, \Big) \Big[
            G(0,1 - v,u)\,H(0,v)
          + G(0,1,1 - v,u)
          + G(0,1,u)\,i\,\pi
          - G(0,1,u)\,H(0,v)\\ &
          - G(0,1,u)\,H(1,v)
          \Big]\\
       &- \frac{1}{48\,u\,(1-u)\,(1-v)^2}\,\Big(44 - 158\,v + 184\,v^2 - 70\,v^3 - 
      22\,u + 110\,u\,v - 193\,u\,v^2 + 105\,u\,v^3 \\ &- 22\,u^2 + 70\,u^2\,v - 61\,u^2\,v^2
       + 13\,u^3\,v\, \Big) \,
          H(0,1,v)
          \\
       &+ \frac{1}{48\,u\,(1-u)\,(1-v)^2}\,\Big(44 - 114\,v + 96\,v^2 - 26\,v^3 - 
      22\,u + 110\,u\,v - 127\,u\,v^2 + 39\,u\,v^3 \\ &- 22\,u^2 + 26\,u^2\,v + 5\,u^2\,v^2 - 
      9\,u^3\,v\, \Big) \, 
           H(0,v)\,i\,\pi
          \\
       &- \frac{1}{48\,u\,(1-u)\,(1-v)^2}\,\Big(160 - 454\,v + 428\,v^2 - 134\,v^3
       - 80\,u + 388\,u\,v - 497\,u\,v^2 \\ &+ 189\,u\,v^3 - 80\,u^2 + 134\,u^2\,v - 41\,u^2\,
      v^2 - 13\,u^3\,v\, \Big) \, 
           \zeta_2
         \\
       &- \frac{1}{288\,u\,(1-u)\,(1-v)^2}\,\Big( - 554\,v + 1108\,v^2 - 554\,v^3
       - 132\,u + 384\,u\,v - 1161\,u\,v^2 \\ &+ 909\,u\,v^3 + 494\,u^2\,v - 657\,u^2\,v^2 + 
      132\,u^3 + 31\,u^3\,v\, \Big) \, 
          H(0,v)
          \\
       &+ \frac{1}{4\,u^2}\,\Big(1 - v + u\,v\, \Big) \, \Big[
            \frac{13}{6}\,G(1,u)\,H(0,v)\,i\,\pi
          - \frac{35}{6}\,G(1,u)\,H(0,1,v)
          + \frac{22}{3}\,G(1 - v,u)\,H(0,0,v)\\ &
          - \frac{22}{3}\,G(1,u)\,H(0,0,v)
          + 2\,G(1 - v,1 - v,u)\,\zeta_2
          + 2\,G(1 - v,1 - v,u)\,H(0,v)\,i\,\pi\\ &
          - 2\,G(1 - v,1 - v,u)\,H(0,1,v)
          - 2\,G(1 - v,1 - v,1,1 - v,u)\\ &
          - 2\,G(1 - v,1 - v,1,u)\,i\,\pi
          + 2\,G(1 - v,1 - v,1,u)\,H(0,v)
          + 2\,G(1 - v,1 - v,1,u)\,H(1,v)\\ &
          + 2\,G(1 - v,0,1 - v,u)\,H(0,v)
          + 2\,G(1 - v,0,1,1 - v,u)
          + 2\,G(1 - v,0,1,u)\,i\,\pi\\ &
          - 2\,G(1 - v,0,1,u)\,H(0,v)
          - 2\,G(1 - v,0,1,u)\,H(1,v)
          + G(1 - v,1,1 - v,u)\,H(0,v)\\ &
          - 4\,G(1 - v,1,u)\,\zeta_2
          - G(1 - v,1,u)\,H(0,v)\,i\,\pi
          + G(1 - v,1,u)\,H(0,1,v)\\ &
          - G(1 - v,1,u)\,H(1,0,v)
          + 2\,G(1 - v,1,1,1 - v,u)
          + 2\,G(1 - v,1,1,u)\,i\,\pi\\ &
          - 2\,G(1 - v,1,1,u)\,H(0,v)
          - 2\,G(1 - v,1,1,u)\,H(1,v)
          + 2\,G(1,1 - v,1 - v,u)\,H(0,v)\\ &
          - 4\,G(1,1 - v,u)\,\zeta_2
          - G(1,1 - v,u)\,H(0,v)\,i\,\pi
          + G(1,1 - v,u)\,H(0,1,v)\\ &
          - G(1,1 - v,u)\,H(1,0,v)
          + 3\,G(1,1 - v,1,1 - v,u)
          + 3\,G(1,1 - v,1,u)\,i\,\pi\\ &
          - 3\,G(1,1 - v,1,u)\,H(0,v)
          - 3\,G(1,1 - v,1,u)\,H(1,v)
          - 3\,G(1,u)\,\zeta_3
          - G(1,u)\,i\,\pi\,\zeta_2\\ &
          + 4\,G(1,u)\,H(1,v)\,\zeta_2
          + G(1,u)\,H(1,0,v)\,i\,\pi
          - G(1,u)\,H(1,0,1,v)\\ &
          + G(1,u)\,H(1,1,0,v)
          - G(1,0,1 - v,u)\,H(0,v)
          - G(1,0,1,1 - v,u)
          - G(1,0,1,u)\,i\,\pi\\ &
          + G(1,0,1,u)\,H(0,v)
          + G(1,0,1,u)\,H(1,v)
          + 2\,G(1,1,1 - v,1 - v,u)\\ &
          + 2\,G(1,1,1 - v,u)\,i\,\pi
          - 2\,G(1,1,1 - v,u)\,H(0,v)
          - 2\,G(1,1,1 - v,u)\,H(1,v)
          - 2\,G(1,1,u)\,\zeta_2\\ &
          - 2\,G(1,1,u)\,H(1,v)\,i\,\pi
          + 2\,G(1,1,u)\,H(1,0,v)
          + 2\,G(1,1,u)\,H(1,1,v)\\ &
          - 2\,G(1,1,1,1 - v,u)
          - 2\,G(1,1,1,u)\,i\,\pi
          + 2\,G(1,1,1,u)\,H(0,v)
          + 2\,G(1,1,1,u)\,H(1,v)
          \Big]\\
       &+ \frac{1}{8\,u^2\,(1-u)}\,\Big(3 - 3\,v - u + 2\,u\,v - u^2 + 3\,u^2\,v - 
      u^3\, \Big) \, \Big[
            G(1,1 - v,u)\,i\,\pi
          - G(1,u)\,H(1,v)\,i\,\pi
          \Big]\\
       &+ \frac{1}{24\,u^2\,(1-u)}\,\Big(13 - 13\,v - 19\,u + 38\,u\,v + 3\,u^2 - 31
      \,u^2\,v + 3\,u^3\, \Big) \, \Big[\\ &
          - G(1 - v,1 - v,u)\,H(0,v)
          - G(1,1 - v,1 - v,u)
          + G(1,1 - v,u)\,H(1,v)
          - G(1,u)\,H(1,1,v)
          \Big]\\
       &+ \frac{1}{24\,u^2\,(1-u)}\,\Big(13 - 13\,v - 7\,u + 14\,u\,v - 3\,u^2 + 5\,
      u^2\,v - 3\,u^3\, \Big) \, \Big[
          - G(1,1,1 - v,u)\\ &
          - G(1,1,u)\,i\,\pi
          + G(1,1,u)\,H(0,v)
          + G(1,1,u)\,H(1,v)
          \Big]\\
       &+ \frac{1}{24\,u^2\,(1-u)}\,\Big(22 - 22\,v - 28\,u + 56\,u\,v + 3\,u^2 - 40
      \,u^2\,v + 3\,u^3\, \Big) \, \Big[
            G(1,1 - v,u)\,H(0,v)\\ &
          - G(1,u)\,H(1,0,v)
          \Big]\\
       &- \frac{1}{24\,u^2\,(1-u)}\,\Big(67 - 67\,v - 61\,u + 122\,u\,v - 3\,u^2 - 
      49\,u^2\,v - 3\,u^3\, \Big) \, 
           G(1,u)\,\zeta_2
          \\
       &+ \frac{1}{144\,u^2\,(1-u)\,(1-v)}\,\Big(277 - 554\,v + 277\,v^2 - 355\,u
       + 969\,u\,v - 614\,u\,v^2 + 21\,u^2 \\ &- 379\,u^2\,v + 340\,u^2\,v^2 + 75\,u^3 - 39\,
      u^3\,v - 18\,u^4\, \Big) \, \Big[
            G(1,1 - v,u)
          + G(1,u)\,i\,\pi\\ &
          - G(1,u)\,H(0,v)
          - G(1,u)\,H(1,v)
          \Big]\\
       &+ \frac{1}{8\,u^2\,(1-u)\,(1-v)^2}\,\Big(3 - 9\,v + 9\,v^2 - 3\,v^3 - 3\,u
       + 14\,u\,v - 19\,u\,v^2 + 8\,u\,v^3 - 3\,u^2\,v \\ &+ 9\,u^2\,v^2 - 6\,u^2\,v^3 - 2\,u^3
      \,v + 3\,u^3\,v^2 - u^4\,v\, \Big) \, 
            G(1 - v,u)\,H(0,v)\,i\,\pi \\ &
        + \frac{1}{24\,u^2\,(1-u)\,(1-v)^2}\,\Big(13 - 39\,v + 39\,v^2 - 13\,v^3 - 
      13\,u + 46\,u\,v - 53\,u\,v^2 + 20\,u\,v^3 \\ &- 13\,u^2\,v + 17\,u^2\,v^2 - 4\,u^2\,v^3
       + 6\,u^3\,v - 9\,u^3\,v^2 + 3\,u^4\,v\, \Big) \, 
           G(1 - v,u)\,H(0,1,v)
          \\
       &- \frac{1}{24\,u^2\,(1-u)\,(1-v)^2}\,\Big(13 - 39\,v + 39\,v^2 - 13\,v^3 + 
      5\,u - 8\,u\,v + u\,v^2 + 2\,u\,v^3 - 9\,u^2 \\ &+ 32\,u^2\,v - 46\,u^2\,v^2 + 23\,u^2\,
      v^3 - 9\,u^3 + 24\,u^3\,v - 18\,u^3\,v^2 + 3\,u^4\,v\, \Big) \, 
           G(1 - v,u)\,\zeta_2
          \\
       &+ \frac{1}{144\,u^2\,(1-u)\,(1-v)^2}\,\Big(277 - 831\,v + 831\,v^2 - 277\,
      v^3 - 223\,u + 850\,u\,v - 1031\,u\,v^2 \\ &+ 404\,u\,v^3 - 45\,u^2 - 70\,u^2\,v + 140
      \,u^2\,v^2 - 25\,u^2\,v^3 + 9\,u^3 + 96\,u^3\,v - 144\,u^3\,v^2 - 18\,u^4 \\ &+ 57\,u^4
      \,v\, \Big) \, 
           G(1 - v,u)\,H(0,v)\, ,
\end{align*}
\begin{align}
   A_\beta^{(3)} &=
          \frac{u}{(1-v)} \, \Bigg[
          + \frac{5}{12}\,\zeta_2
          + \frac{11}{16}\,i\,\pi\,\zeta_2
          + \frac{11}{16}\,H(0,v)\,\zeta_2
          - \frac{49}{32}\,\zeta_4
          - \frac{7}{4}\,i\,\pi\,\zeta_3
          - \frac{7}{4}\,H(0,v)\,\zeta_3
          - \frac{389}{144}\,\zeta_3\hspace{10mm}\nonumber \\ &
          + \frac{2759}{864}\,i\,\pi
          + \frac{2759}{864}\,H(0,v)
          + \frac{81659}{10368}
          \Bigg]\, ,\nonumber \\ \nonumber \\
   A_\beta^{(4)} &= 0 \, , 
\end{align}

\begin{align*}
   A_\gamma^{(1)} &=
          \frac{v\,(1-u-v)}{8\,u\,(1-u)^3}\,\Big(2 - 5\,u + 5\,u^2\, \Big) \, \Big[
          - 2\,i\,\pi\,\zeta_2
          - 2\,G(0,u)\,\zeta_2
          - G(0,u)\,H(1,0,v)\\ &
          - H(1,v)\,\zeta_2
          - H(1,0,v)\,i\,\pi
          - H(1,1,0,v)
          \Big]\\
       &+ \frac{v\,(1-u-v)}{4\,u^2} \, \Big[
            G(1 - v,1 - v, - v,u)\,i\,\pi
          - G(1 - v,1 - v, - v,u)\,H(0,v)\\ &
          + G(1 - v,1 - v, - v,0,u)
          + G(1 - v,1 - v,u)\,\zeta_2
          + G(1 - v,1 - v,u)\,H(0,v)\,i\,\pi\\ &
          - G(1 - v,1 - v,u)\,H(0,0,v)
          + G(1 - v,1 - v,u)\,H(1,0,v)
          + G(1 - v,1 - v,0,u)\,H(0,v)\\ &
          + G(1 - v,u)\,\zeta_3
          + G(1 - v,u)\,H(0,v)\,\zeta_2
          + G(1 - v,u)\,H(0,0,v)\,i\,\pi\\ &
          - G(1 - v,u)\,H(0,0,0,v)
          + G(1 - v,u)\,H(0,1,0,v)
          + G(1 - v,u)\,H(1,0,0,v)\\ &
          - G(1 - v,0, - v,u)\,i\,\pi
          + G(1 - v,0, - v,u)\,H(0,v)
          - G(1 - v,0, - v,0,u)\\ &
          + G(1 - v,0,u)\,H(0,0,v)
          + G(1,1 - v, - v,u)\,i\,\pi
          - G(1,1 - v, - v,u)\,H(0,v)\\ &
          + G(1,1 - v, - v,0,u)
          + G(1,1 - v,u)\,\zeta_2
          + G(1,1 - v,u)\,H(0,v)\,i\,\pi\\ &
          - G(1,1 - v,u)\,H(0,0,v)
          + G(1,1 - v,u)\,H(1,0,v)
          + G(1,1 - v,0,u)\,H(0,v)\\ &
          + G(1,u)\,\zeta_3
          - 2\,G(1,u)\,i\,\pi\,\zeta_2
          - G(1,u)\,H(0,v)\,\zeta_2
          - G(1,u)\,H(0,0,v)\,i\,\pi\\ &
          + G(1,u)\,H(0,0,0,v)
          - G(1,u)\,H(0,1,0,v)
          - G(1,u)\,H(1,v)\,\zeta_2
          - G(1,u)\,H(1,0,v)\,i\,\pi\\ &
          + G(1,u)\,H(1,0,0,v)
          - G(1,u)\,H(1,1,0,v)
          + G(1,0, - v,u)\,i\,\pi
          - G(1,0, - v,u)\,H(0,v)\\ &
          + G(1,0, - v,0,u)
          - 2\,G(1,0,u)\,\zeta_2
          - G(1,0,u)\,H(0,0,v)
          - G(1,0,u)\,H(1,0,v)
          \Big]\\
       &+ \frac{v}{8\,u\,(1-u)^3\,(1-v)^2}\,\Big(4 - 10\,v + 8\,v^2 - 2\,v^3 - 12\,u
       + 30\,u\,v - 23\,u\,v^2 + 5\,u\,v^3 + 12\,u^2 \\ &- 34\,u^2\,v + 26\,u^2\,v^2 - 5\,u^2\,
      v^3 - u^3 + 11\,u^3\,v - 7\,u^3\,v^2 - 4\,u^4 + u^4\,v + u^5\, \Big) \, \Big[
            \zeta_3\\ &
          + G(1 - v, - v,u)\,i\,\pi
          - G(1 - v, - v,u)\,H(0,v)
          + G(1 - v, - v,0,u)
          + G(1 - v,u)\,\zeta_2\\ &
          + G(1 - v,u)\,H(0,v)\,i\,\pi
          - G(1 - v,u)\,H(0,0,v)
          + G(1 - v,u)\,H(1,0,v)\\ &
          + G(1 - v,0,u)\,H(0,v)
          + H(1,0,0,v)
          \Big]\\
       &+ \frac{v}{8\,u\,(1-u)^3\,(1-v)^2}\,\Big(2\,v - 4\,v^2 + 2\,v^3 + 2\,u - 8\,u
      \,v + 11\,u\,v^2 - 5\,u\,v^3 - 8\,u^2 + 16\,u^2\,v \\ &- 14\,u^2\,v^2 + 5\,u^2\,v^3 + 9\,
      u^3 - 9\,u^3\,v + 3\,u^3\,v^2 - 4\,u^4 + u^4\,v + u^5\, \Big) \, \Big[
          - G(0, - v,u)\,i\,\pi\\ &
          + G(0, - v,u)\,H(0,v)
          - G(0, - v,0,u)
          + G(0,u)\,H(0,0,v)
          + H(0,v)\,\zeta_2
          + H(0,0,v)\,i\,\pi\\ &
          - H(0,0,0,v)
          + H(0,1,0,v)
          \Big]\\
       &+ \frac{(1-u-v)}{288\,(1-v)^2}\,\Big(132 + 31\,v\, \Big) \, 
           H(0,v)
        + \frac{(1-u-v)}{16\,(1-u)\,(1-v)}\,\Big(3 - 2\,v - u\, \Big) \, 
           i\,\pi
          \\
       &- \frac{(1-u-v)}{48\,(1-u)\,(1-v)}\,\Big(13 + 6\,v - 19\,u\, \Big) \, 
           G(0,u)
          \\
       &+ \frac{1}{16\,(1-u)^2\,(1-v)^2}\,\Big(2 + 3\,v - 7\,v^2 + 2\,v^3 - 4\,u - 
      13\,u\,v + 20\,u\,v^2 - 6\,u\,v^3 + 4\,u^2 + 7\,u^2\,v \\ &- 5\,u^2\,v^2 - 2\,u^3 - u^3\,
      v\, \Big) \, 
           H(0,v)\,i\,\pi
          \\
       &+ \frac{1}{48\,(1-u)^2\,(1-v)^2}\,\Big(6 - 13\,v + v^2 + 6\,v^3 - 12\,u + 
      27\,u\,v + 16\,u\,v^2 - 18\,u\,v^3 + 12\,u^2 \\ &- 45\,u^2\,v + 7\,u^2\,v^2 - 6\,u^3 + 
      19\,u^3\,v\, \Big) \, 
           G(0,u)\,H(0,v)
          \\
       &- \frac{1}{288\,u\,(1-v)}\,\Big(144\,v - 144\,v^2 - 133\,u + 133\,u\,v + 277
      \,u^2\, \Big) \, \\
       &+ \frac{1}{8\,u\,(1-u)^2\,(1-v)}\,\Big( - 2\,v^2 + 2\,v^3 - u - u\,v + 7\,u\,
      v^2 - 4\,u\,v^3 + 2\,u^2 + 2\,u^2\,v - 5\,u^2\,v^2 \\ &- 2\,u^3 + u^3\,v + u^4\, \Big)
       \, \Big[
          - G( - v,u)\,i\,\pi
          + G( - v,u)\,H(0,v)
          - G( - v,0,u)
          \Big]\\
       &- \frac{1}{24\,u\,(1-u)^2\,(1-v)^2}\,\Big(6\,v^2 - 12\,v^3 + 6\,v^4 + 3\,u
       - 22\,u\,v - 2\,u\,v^2 + 33\,u\,v^3 - 12\,u\,v^4 - 6\,u^2 \\ &+ 66\,u^2\,v - 23\,u^2\,
      v^2 - 15\,u^2\,v^3 + 6\,u^3 - 75\,u^3\,v + 25\,u^3\,v^2 - 3\,u^4 + 25\,u^4\,v \, \Big)
        \, 
           H(0,0,v)
          \\
       &+ \frac{1}{48\,u\,(1-u)^2\,(1-v)^2}\,\Big(12\,v - 36\,v^2 + 36\,v^3 - 12\,
      v^4 + 6\,u - 49\,u\,v + 97\,u\,v^2 - 78\,u\,v^3 + 24\,u\,v^4 \\ &- 12\,u^2 + 51\,u^2\,v
       - 32\,u^2\,v^2 + 6\,u^2\,v^3 + 12\,u^3 - 45\,u^3\,v + 7\,u^3\,v^2 - 6\,u^4 + 19\,
      u^4\,v\, \Big) \, 
           H(1,0,v)
          \\
       &+ \frac{1}{48\,u\,(1-u)^2\,(1-v)^2}\,\Big(24\,v - 72\,v^2 + 72\,v^3 - 24\,
      v^4 + 6\,u - 85\,u\,v + 193\,u\,v^2 - 162\,u\,v^3 \\ &+ 48\,u\,v^4 - 12\,u^2 + 75\,u^2\,
      v - 80\,u^2\,v^2 + 30\,u^2\,v^3 + 12\,u^3 - 45\,u^3\,v + 7\,u^3\,v^2 - 6\,u^4 + 19
      \,u^4\,v\, \Big) \, 
           \zeta_2 \, ,\\ \\
   A_\gamma^{(2)} &=
          \frac{v\,(1-u-v)}{8\,u\,(1-u)^2}\,\Big(2 - 3\,u\, \Big) \, \Big[
          - 3\,\zeta_3
          - i\,\pi\,\zeta_2
          - G(1 - v,u)\,H(1,0,v)
          + 4\,H(1,v)\,\zeta_2\\ &
          + H(1,0,v)\,i\,\pi
          - H(1,0,1,v)
          + H(1,1,0,v)
          \Big]\\
       &+ \frac{v\,(1-u-v)}{16\,u\,(1-u)^2}\,\Big(6 - 5\,u\, \Big) \, \Big[
            G(1 - v,u)\,i\,\pi
          - H(1,v)\,i\,\pi
          \Big]\\
       &- \frac{v\,(1-u-v)}{24\,u\,(1-u)^2}\,\Big(22 - 39\,u\, \Big) \, 
          H(1,0,v)
          \\
       &+ \frac{v\,(1-u-v)}{48\,u\,(1-u)^2}\,\Big(26 - 51\,u\, \Big) \, \Big[
          - G(1 - v,1 - v,u)
          + G(1 - v,u)\,H(1,v)
          - H(1,1,v)
          \Big]\\
       &+ \frac{v\,(1-u-v)}{8\,u\,(1-u)^2\,(1-v)^2}\,\Big(4 - 10\,v + 6\,v^2 - 6\,u
       + 14\,u\,v - 9\,u\,v^2 + 2\,u^2\,v - u^3\, \Big) \, \\ & \times \Big[
            G(1 - v,1,1 - v,u)
          + G(1 - v,1,u)\,i\,\pi
          - G(1 - v,1,u)\,H(0,v)
          - G(1 - v,1,u)\,H(1,v)
          \Big]\\
       &+ \frac{v\,(1-u-v)}{8\,u\,(1-u)^2\,(1-v)^2}\,\Big(2\,v - 2\,v^2 - 2\,u\,v + 3
      \,u\,v^2 - 2\,u^2\,v + u^3\, \Big) \, \Bigg[
            \frac{22}{3}\,H(0,0,v)\\ &
          + G(0,1 - v,u)\,H(0,v)
          + G(0,1,1 - v,u)
          + G(0,1,u)\,i\,\pi
          - G(0,1,u)\,H(0,v)\\ &
          - G(0,1,u)\,H(1,v)
          \Bigg]\\
       &+ \frac{v\,(1-u-v)}{48\,u\,(1-u)^2\,(1-v)^2}\,\Big(44 - 114\,v + 70\,v^2 - 
      66\,u + 158\,u\,v - 105\,u\,v^2 + 26\,u^2\,v - 13\,u^3\, \Big) \,\\ & \times \Big[
          - H(0,1,v)
          \Big]\\
       &+ \frac{v\,(1-u-v)}{48\,u\,(1-u)^2\,(1-v)^2}\,\Big(44 - 70\,v + 26\,v^2 - 
      66\,u + 114\,u\,v - 39\,u\,v^2 - 18\,u^2\,v \\ &+ 9\,u^3\, \Big) \,
          H(0,v)\,i\,\pi
          \\
       &- \frac{v\,(1-u-v)}{48\,u\,(1-u)^2\,(1-v)^2}\,\Big(160 - 294\,v + 134\,v^2
       - 228\,u + 430\,u\,v - 189\,u\,v^2 - 26\,u^2\,v + 13\,u^3\, \Big) \,
           \zeta_2
          \\
       &+ \frac{v\,(1-u-v)}{4\,u^2} \, \Bigg[
            \frac{13}{6}\,G(1,u)\,H(0,v)\,i\,\pi
          - \frac{35}{6}\,G(1,u)\,H(0,1,v)
          + \frac{22}{3}\,G(1 - v,u)\,H(0,0,v)\\ &
          - \frac{22}{3}\,G(1,u)\,H(0,0,v)
          + 2\,G(1 - v,1 - v,u)\,\zeta_2
          + 2\,G(1 - v,1 - v,u)\,H(0,v)\,i\,\pi\\ &
          - 2\,G(1 - v,1 - v,u)\,H(0,1,v)
          - 2\,G(1 - v,1 - v,1,1 - v,u)
          - 2\,G(1 - v,1 - v,1,u)\,i\,\pi\\ &
          + 2\,G(1 - v,1 - v,1,u)\,H(0,v)
          + 2\,G(1 - v,1 - v,1,u)\,H(1,v)
          + 2\,G(1 - v,0,1 - v,u)\,H(0,v)\\ &
          + 2\,G(1 - v,0,1,1 - v,u)
          + 2\,G(1 - v,0,1,u)\,i\,\pi
          - 2\,G(1 - v,0,1,u)\,H(0,v)\\ &
          - 2\,G(1 - v,0,1,u)\,H(1,v)
          + G(1 - v,1,1 - v,u)\,H(0,v)
          - 4\,G(1 - v,1,u)\,\zeta_2\\ &
          - G(1 - v,1,u)\,H(0,v)\,i\,\pi
          + G(1 - v,1,u)\,H(0,1,v)
          - G(1 - v,1,u)\,H(1,0,v)\\ &
          + 2\,G(1 - v,1,1,1 - v,u)
          + 2\,G(1 - v,1,1,u)\,i\,\pi
          - 2\,G(1 - v,1,1,u)\,H(0,v)\\ &
          - 2\,G(1 - v,1,1,u)\,H(1,v)
          + 2\,G(1,1 - v,1 - v,u)\,H(0,v)
          - 4\,G(1,1 - v,u)\,\zeta_2\\ &
          - G(1,1 - v,u)\,H(0,v)\,i\,\pi
          + G(1,1 - v,u)\,H(0,1,v)
          - G(1,1 - v,u)\,H(1,0,v)\\ &
          + 3\,G(1,1 - v,1,1 - v,u)
          + 3\,G(1,1 - v,1,u)\,i\,\pi
          - 3\,G(1,1 - v,1,u)\,H(0,v)\\ &
          - 3\,G(1,1 - v,1,u)\,H(1,v)
          - 3\,G(1,u)\,\zeta_3
          - G(1,u)\,i\,\pi\,\zeta_2
          + 4\,G(1,u)\,H(1,v)\,\zeta_2\\ &
          + G(1,u)\,H(1,0,v)\,i\,\pi
          - G(1,u)\,H(1,0,1,v)
          + G(1,u)\,H(1,1,0,v)\\ &
          - G(1,0,1 - v,u)\,H(0,v)
          - G(1,0,1,1 - v,u)
          - G(1,0,1,u)\,i\,\pi
          + G(1,0,1,u)\,H(0,v)\\ &
          + G(1,0,1,u)\,H(1,v)
          + 2\,G(1,1,1 - v,1 - v,u)
          + 2\,G(1,1,1 - v,u)\,i\,\pi\\ &
          - 2\,G(1,1,1 - v,u)\,H(0,v)
          - 2\,G(1,1,1 - v,u)\,H(1,v)
          - 2\,G(1,1,u)\,\zeta_2\\ &
          - 2\,G(1,1,u)\,H(1,v)\,i\,\pi
          + 2\,G(1,1,u)\,H(1,0,v)
          + 2\,G(1,1,u)\,H(1,1,v)\\ &
          - 2\,G(1,1,1,1 - v,u)
          - 2\,G(1,1,1,u)\,i\,\pi
          + 2\,G(1,1,1,u)\,H(0,v)
          + 2\,G(1,1,1,u)\,H(1,v)
          \Bigg]\\
       &+ \frac{v\,(1-u-v)}{8\,u^2\,(1-u)^2}\,\Big(3 - 2\,u - 3\,u^2\, \Big) \, \Big[
            G(1,1 - v,u)\,i\,\pi
          - G(1,u)\,H(1,v)\,i\,\pi
          \Big]\\
       &+ \frac{v\,(1-u-v)}{12\,u^2\,(1-u)^2}\,\Big(11 - 28\,u + 20\,u^2\, \Big)
       \, \Big[
            G(1,1 - v,u)\,H(0,v)
          - G(1,u)\,H(1,0,v)
          \Big]\\
       &+ \frac{v\,(1-u-v)}{24\,u^2\,(1-u)^2}\,\Big(13 - 38\,u + 31\,u^2\, \Big)
       \, \Big[
          - G(1 - v,1 - v,u)\,H(0,v)
          - G(1,1 - v,1 - v,u)\\ &
          + G(1,1 - v,u)\,H(1,v)
          - G(1,u)\,H(1,1,v)
          \Big]\\
       &+ \frac{v\,(1-u-v)}{24\,u^2\,(1-u)^2}\,\Big(13 - 14\,u - 5\,u^2\, \Big) \, \Big[
          - G(1,1,1 - v,u)
          - G(1,1,u)\,i\,\pi
          + G(1,1,u)\,H(0,v)\\ &
          + G(1,1,u)\,H(1,v)
          \Big]\\
       &- \frac{v\,(1-u-v)}{24\,u^2\,(1-u)^2}\,\Big(67 - 122\,u + 49\,u^2\, \Big)
       \, 
           G(1,u)\,\zeta_2
          \\
       &+ \frac{v\,(1-u-v)}{8\,u^2\,(1-u)^2\,(1-v)^2}\,\Big(3 - 6\,v + 3\,v^2 - 6\,u
       + 14\,u\,v - 8\,u\,v^2 + 3\,u^2 - 8\,u^2\,v + 6\,u^2\,v^2 \\ &- 2\,u^3\,v + u^4\, \Big)
       \, 
           G(1 - v,u)\,H(0,v)\,i\,\pi
          \\
       &+ \frac{v\,(1-u-v)}{24\,u^2\,(1-u)^2\,(1-v)^2}\,\Big(13 - 26\,v + 13\,v^2
       - 26\,u + 46\,u\,v - 20\,u\,v^2 + 13\,u^2 - 20\,u^2\,v \\ &+ 4\,u^2\,v^2 + 6\,u^3\,v - 
      3\,u^4\, \Big) \, 
           G(1 - v,u)\,H(0,1,v)
          \\
       &- \frac{v\,(1-u-v)}{24\,u^2\,(1-u)^2\,(1-v)^2}\,\Big(13 - 26\,v + 13\,v^2
       - 8\,u + 10\,u\,v - 2\,u\,v^2 - 14\,u^2 + 34\,u^2\,v \\ &- 23\,u^2\,v^2 + 6\,u^3\,v - 3
      \,u^4\, \Big) \, 
          G(1 - v,u)\,\zeta_2
          \\
       &+ \frac{(1-u-v)}{288\,u\,(1-u)^2\,(1-v)}\,\Big( - 554\,v + 554\,v^2 + 36\,u
       + 987\,u\,v - 909\,u\,v^2 - 150\,u^2 - 78\,u^2\,v \\ &+ 114\,u^3\, \Big) \, \Big[
          - G(1 - v,u)
          + H(1,v)
          \Big]\\
       &+ \frac{(1-u-v)}{288\,u\,(1-u)^2\,(1-v)}\,\Big(554\,v - 554\,v^2 - 36\,u - 
      855\,u\,v + 909\,u\,v^2 + 18\,u^2 - 54\,u^2\,v + 18\,u^3\, \Big) \,
           i\,\pi
          \\
       &+ \frac{(1-u-v)}{288\,u\,(1-u)^2\,(1-v)^2}\,\Big(554\,v^2 - 554\,v^3 + 96\,
      u\,v - 842\,u\,v^2 + 909\,u\,v^3 - 132\,u^2 + 228\,u^2\,v \\ &- 422\,u^2\,v^2 + 132\,
      u^3 + 31\,u^3\,v\, \Big) \,
           H(0,v)
          \\
       &+ \frac{(1-u-v)}{144\,u^2\,(1-u)^2\,(1-v)}\,\Big(277\,v - 277\,v^2 - 18\,u
       - 614\,u\,v + 614\,u\,v^2 + 36\,u^2 + 358\,u^2\,v \\ &- 340\,u^2\,v^2 - 36\,u^3 - 18\,
      u^3\,v + 18\,u^4\, \Big) \, \Big[
            G(1,1 - v,u)
          + G(1,u)\,i\,\pi
          - G(1,u)\,H(0,v)\\ &
          - G(1,u)\,H(1,v)
          \Big]\\
       &+ \frac{(1-u-v)}{144\,u^2\,(1-u)^2\,(1-v)^2}\,\Big( - 277\,v + 554\,v^2 - 
      277\,v^3 + 18\,u + 464\,u\,v - 886\,u\,v^2 + 404\,u\,v^3 \\ &- 36\,u^2 - 124\,u^2\,v + 
      224\,u^2\,v^2 - 25\,u^2\,v^3 + 36\,u^3 - 18\,u^3\,v - 96\,u^3\,v^2 - 18\,u^4 + 57\,
      u^4\,v\, \Big) \, \\ & \times \Big[
          - G(1 - v,u)\,H(0,v)
          \Big]\\
       &+ \frac{1}{288\,u\,(1-u)\,(1-v)}\,\Big(144\,v - 144\,v^2 + 144\,u - 11\,u\,v
       - 133\,u\,v^2 - 421\,u^2 + 144\,u^2\,v + 277\,u^3\, \Big) \,, 
\end{align*}
\begin{align}
   A_\gamma^{(3)} &= 0 \,,\hspace{15cm}\nonumber \\ \nonumber \\
   A_\gamma^{(4)} &= 0 \,.
\end{align}
}

\section{Helicity amplitudes for $q\bar q\to Vg$ and $qg \to Vq$}
\label{app:vjet}
In~\cite{3jtensor} the helicity amplitudes for the processes
\begin{equation}
\gamma^* (p_4) \longrightarrow q(p_1) + \bar q (p_2) + g(p_3)\;
\end{equation}
have been computed. As already discussed, 
the kinematical region relevant for $3$-jet production
is characterised by $q^2$ and $s_{ij}$ all positive. 

Following~\cite{ancont}, we can analytically continue these matrix elements
to the kinematical configuration relevant for $(V + 1j)$ production at
hadron colliders, which is the same kinematical situation relevant for 
the $V \gamma$ production studied above:
\begin{equation}
  q(p_2) + \bar{q}(p_1) \longrightarrow g(-p_3) + \gamma^*(p_4)\;, \label{qqbgV}
\end{equation}
and
\begin{equation}
 q(p_2) + g(p_3) \longrightarrow q(-p_1) + \gamma^*(p_4)\;, \label{qgqV}
\end{equation}
The kinematical situation in~\eqref{qqbgV} is the same as for,
\begin{equation}
q(p_2) + \bar{q}(p_1) \longrightarrow \gamma(-p_3) + V(p_4)\;,
\end{equation}
so we can introduce again the dimensionless variables
\begin{equation} 
u = -\frac{s_{13}}{s_{12}}=-\frac{y}{x}\,, \qquad v = \frac{q^2}{s_{12}} =
 \frac{1}{x} \;,
\end{equation}
which fulfil
\begin{displaymath}
0\leq u \leq v\,, \qquad 0\leq v \leq 1\;.
\end{displaymath}
On the other hand, the kinematical situation in~\eqref{qgqV} can be described
with the following choice of dimensionless variables~\cite{ancont}:
\begin{equation} 
u'  = -\frac{s_{13}}{s_{23}}=-\frac{y}{z}\,, \qquad v' = \frac{q^2}{s_{23}} =
 \frac{1}{z} \;,
\end{equation}
which fulfil again:
\begin{displaymath}
0\leq u' \leq v'\,, \qquad 0\leq v' \leq 1\;.
\end{displaymath}

The helicity amplitude coefficients $\alpha$,
$\beta$ and $\gamma$ are vectors in colour space and
have perturbative expansions:
\begin{equation}
\Omega =  \sqrt{4\pi\alpha} \sqrt{4\pi\alpha_s} \; \bom{T}^a_{ij}\, \left[
\Omega^{(0)}  
+ \left(\frac{\alpha_s}{2\pi}\right) \Omega^{(1)}  
+ \left(\frac{\alpha_s}{2\pi}\right)^2 \Omega^{(2)} 
+ {\cal O}(\alpha_s^3) \right] \;,\nonumber \\
\end{equation}
for $\Omega = \alpha,\beta,\gamma$. The dependence on $(u,v)$ 
or $(u',v')$ is again 
implicit.

The ultraviolet and infrared properties of the helicity coefficients are
fully described in~\cite{3jtensor}, 
\begin{eqnarray}
\Omega^{(0)}  &=& \Omega^{(0),{\rm un}} ,
 \nonumber \\
\Omega^{(1)}  &=& 
S_\e^{-1} \Omega^{(1),{\rm un}} 
-\frac{\beta_0}{2\e} \Omega^{(0),{\rm un}}  ,  \nonumber \\
\Omega^{(2)} &=& 
S_\e^{-2} \Omega^{(2),{\rm un}}  
-\frac{3\beta_0}{2\e} S_\e^{-1}
\Omega^{(1),{\rm un}}  
-\left(\frac{\beta_1}{4\e}-\frac{3\beta_0^2}{8\e^2}\right)
\Omega^{(0),{\rm un}},
\end{eqnarray}
and
\begin{eqnarray}
\Omega^{(1)} &=& {\bom I}^{(1)}(\epsilon) \Omega^{(0)} +
\Omega^{(1),{\rm finite}},\nonumber \\
\Omega^{(2)} &=& \Biggl (-\frac{1}{2}  {\bom I}^{(1)}(\epsilon) {\bom I}^{(1)}(\epsilon)
-\frac{\beta_0}{\epsilon} {\bom I}^{(1)}(\epsilon) 
+e^{-\epsilon \gamma } \frac{ \Gamma(1-2\epsilon)}{\Gamma(1-\epsilon)} 
\left(\frac{\beta_0}{\epsilon} + K\right)
{\bom I}^{(1)}(2\epsilon) + {\bom H}^{(2)}(\epsilon) 
\Biggr )\Omega^{(0)}\nonumber \\
&& + {\bom I}^{(1)}(\epsilon) \Omega^{(1)}+ \Omega^{(2),{\rm finite}},
\end{eqnarray}
where the infrared singularity operator $\bom{I}^{(1)}(\epsilon)$ is a $1 \times 1$ matrix 
in colour space and is given by
\begin{equation}
\bom{I}^{(1)}(\epsilon)
=
- \frac{e^{\epsilon\gamma}}{2\Gamma(1-\epsilon)} \Biggl[
N \left(\frac{1}{\epsilon^2}+\frac{3}{4\epsilon}+\frac{\beta_0}{2N\epsilon}\right) 
\left({\tt S}_{13}+{\tt S}_{23}\right)-\frac{1}{N}
\left(\frac{1}{\epsilon^2}+\frac{3}{2\epsilon}\right)
{\tt S}_{12}\Biggr ]\; ,\label{eq:I1qqg}
\end{equation}
where (since we have set $\mu^2 = s_{123}$)
\begin{equation}
{\tt S}_{ij} = \left(-\frac{s_{123}}{s_{ij}}\right)^{\epsilon}.
\end{equation}
Expanding ${\tt S}_{ij}$,
imaginary parts are generated, the sign of which is fixed by the small imaginary
part $+i0$ of $s_{ij}$.\\
Finally we have
\begin{equation}
\label{eq:htwo}
{\bom H}^{(2)}(\epsilon)
=\frac{e^{\epsilon \gamma}}{4\,\epsilon\,\Gamma(1-\epsilon)} H^{(2)} \;,  
\end{equation}
where, in the \MSbar\ scheme:
\begin{eqnarray}
\label{eq:Htwo}
H^{(2)} &=&  
\left(4\zeta_3+\frac{589}{432}- \frac{11\pi^2}{72}\right)N^2
+\left(-\frac{1}{2}\zeta_3-\frac{41}{54}-\frac{\pi^2}{48} \right)
+\left(-3\zeta_3 -\frac{3}{16} + \frac{\pi^2}{4}\right) \frac{1}{N^2}\nonumber \\
&&
+\left(-\frac{19}{18}+\frac{\pi^2}{36} \right) N\NF 
+\left(-\frac{1}{54}-\frac{\pi^2}{24}\right) \frac{\NF}{N}+ \frac{5}{27} \NF^2.
\end{eqnarray}

They finite one-loop amplitudes can be decomposed 
according to their colour structure as follows:
\begin{equation}
\Omega^{(1),{\rm finite}}(u,v) =  
N\, a_{\Omega}(u,v) + \frac{1}{N}\, b_{\Omega}(u,v) + \beta_0\, c_{\Omega}(u,v)
   \;.
\end{equation}
We attach to the arXiv submission of the paper the one loop coefficients expanded
up to ${\cal O}( \epsilon^2 )$.\newline

The finite two-loop remainder is obtained by subtracting the
predicted infrared structure (expanded through to ${\cal O}(\epsilon^0)$) from
the renormalised helicity coefficient.  We further decompose the 
finite remainder according to the colour structure, as follows:
\begin{align*}
\Omega^{(2),{\rm finite}}(u,v) &=  
N^2 A_\Omega(u,v) + B_\Omega(u,v) + \frac{1}{N^2} C_\Omega(u,v) 
+ N\NF D_\Omega(u,v) \\ 
&+ \frac{\NF}{N} E_\Omega(u,v) + \NF^2 F_\Omega (u,v)
 + \NFZ \left(\frac{4}{N}-N\right) G_\Omega(u,v) \; , 
\label{eq:twoloopamp}
\end{align*}
The complete expressions in FORM format can be found 
attached to the arXiv submission of this paper.

\end{document}